\documentclass[]{aastex62}

\usepackage[utf8]{inputenc}
\usepackage[version=4]{mhchem} 
\usepackage{amsmath}
\usepackage{amssymb}
\usepackage{graphics}
\usepackage{graphicx}
\usepackage{rotating}
\usepackage{color}
\usepackage{epstopdf}
\usepackage{orcidlink}

\usepackage{ulem}

\graphicspath{{./}{figures/}}

\newcommand{\mj}[1]{{\color{magenta}#1}}

\begin{document}

\title{The robust detection and spatial distribution of acetaldehyde in Orion KL: ALMA observations and chemical modeling}

\author{Miwha Jin}
\affil{Astrochemistry Laboratory, Code 691, NASA Goddard Space Flight Center, Greenbelt, MD 20771, USA}
\affil{Department of Physics, Catholic University of America, Washington, DC 20064, USA}

\author{Anthony J. Remijan}
\affiliation{National Radio Astronomy Observatory, Charlottesville, VA, 22903, USA}
\affiliation{Department of Astronomy, University of Virginia, Charlottesville, VA, 22904, USA}

\author{Robin T. Garrod}
\affiliation{Department of Chemistry, University of Virginia, Charlottesville, VA, 22904, USA}
\affiliation{Department of Astronomy, University of Virginia, Charlottesville, VA, 22904, USA}

\author{Giseon Baek}
\affiliation{Department of Physics and Astronomy, Seoul National University, 1 Gwanak-ro, Gwanak-gu, Seoul 08826, Republic of Korea}
\affiliation{Research Institute of Basic Sciences, Seoul National University, Seoul 08826, Republic of Korea}

\author{Martin Cordiner}
\affiliation{Astrochemistry Laboratory, Code 691, NASA Goddard Space Flight Center, Greenbelt, MD 20771, USA}
\affiliation{Department of Physics, Catholic University of America, Washington, DC 20064, USA}

\author{Steven Charnley}
\affiliation{Astrochemistry Laboratory, Code 691, NASA Goddard Space Flight Center, Greenbelt, MD 20771, USA}

\author{Eric Herbst}
\affiliation{Department of Chemistry, University of Virginia, Charlottesville, VA, 22904, USA}
\affiliation{Department of Astronomy, University of Virginia, Charlottesville, VA, 22904, USA}

\author{Jeong-Eun Lee}
\affiliation{Department of Physics and Astronomy, Seoul National University, 1 Gwanak-ro, Gwanak-gu, Seoul 08826, Republic of Korea}
\affiliation{SNU Astronomy Research Center, Seoul National University, 1 Gwanak-ro, Gwanak-gu, Seoul 08826, Republic of Korea}

\begin{abstract}
Despite the organic molecule inventory detected in Orion KL, acetaldehyde ($\ce{CH3CHO}$) -- one of the most ubiquitous interstellar aldehydes -- has not been firmly identified with mm-wave interferometry. We analyze extensive ALMA archival datasets (142-355 GHz) to search for acetaldehyde, revealing two distinct acetaldehyde emission peaks and one component with more complex kinematic structures. One peak aligns with MF10/IRc2, where emissions of other O-bearing complex organic molecules are rarely reported, while the other is coincident with the ethanol peak in the hot core-SW. The MF10/IRc2 detection suggests unique chemistry, possibly influenced by repeated heating events. In contrast, co-detection with ethanol indicates an ice origin and suggests a potential chemical relationship between the two species. We determined acetaldehyde column densities and its kinetic temperatures toward these two peaks under LTE assumptions and compare its distribution with ethanol and other molecules that have an aldehyde (HCO) group, such as methyl formate, glycolaldehyde, and formic acid. Toward the ethanol peak, observed abundance ratios between HCO-containing species are analyzed using a chemical model. The model suggests two key points: 1) the destruction of ethanol to form acetaldehyde in the ice may contribute to the observed correlation between the two species, 2) a long cold-collapse timescale and a methyl formate binding energy similar to or lower than water are needed to explain the observations. The relative abundance ratios obtained from the model are highly sensitive to the assumed kinetic temperature, which accounts for the high spatial variability of the aldehyde ratios observed toward Orion KL.
\end{abstract}

\keywords{
Astrochemistry (75) --
Star formation (1569) --
Complex organic molecules (2256) --
Observational astronomy(1145)
}

\section{Introduction} \label{sec:intro}

The Orion molecular cloud complex is one of the most well-studied high-mass star forming regions due to its proximity \citep[388 pc $\pm$ 5 pc;][]{kounkel17}. It exhibits a highly structured hierarchy of star-forming regions, hosting many substructures such as Orion molecular clouds (OMCs) 1-4. The Orion Kleinmann-Low Nebula (Orion KL) in OMC-1 has received particular attention because of its unique features. The enormously high IR luminosity \citep[$\thicksim10^5~L_{\odot}$][]{winn-williams84} and chemical complexity on various scales~\citep{blake87, blake96, liu02, friedelandsnyder08, favre11, pagani17, tercero18, peng19} are attributed to a recent ($\thicksim$500 yr) explosive event~\citep{zapata11}; investigations of the proper motions of infrared/radio sources BN, I, and n  \citep{mentenandreid95, lonsdale82} reveal that they are moving away from a common central region~\citep{gomez05}. Orion KL thus provides a unique laboratory for studying a rich interstellar chemistry driven by a very recent and energetic event at the heart of the nebula.

Orion KL is classically described as being host to four large-scale components that show a distinctive systemic velocity; these are the hot core (HC), the compact ridge (CR), the extended ridge, and the plateau~\citep{blake87}. In particular, HC and CR are known as major emission regions of organic molecules.
Observations of O- and N-bearing species toward these two regions show apparent chemical segregation; N-bearing molecules tend to peak north of HC, whereas O-bearing molecules are found in both HC and CR~\citep{guelin08, beuther05, friedelandsnyder08}. The HC is believed to be externally heated; its heating source is possibly either an impact of the explosive event or the shocks from the source I outflow. The CR would also be externally heated, but is not directly affected by the impact of the explosion~\citep{pagani17}.

Complex organic molecules (COMs) are typically defined as carbon-bearing molecules composed of six or more atoms.
The high angular resolution achieved by interferometric observations allows us to understand the distributions of many COMs in  Orion KL. For example, many methyl formate (\ce{CH3OCHO}; MF) peaks are identified~\citep{favre11}, and \citet{tercero15} provide maps of species containing the functional groups formate, alcohol, and ether, including both the methyl and ethyl derivatives. Additionally, \citet[][TCL18]{tercero18} reported a clear segregation of complex O-bearing species depending on their different functional groups. Deeper observations with interferometry have also revealed many new species not previously detected in Orion KL. For example, \citet{pagani17} present the detection of new species such as propyl cyanide (\ce{C3H7CN}), while \citet{favre17} reported the first detection of gGg' ethylene glycol ($\ce{(CH2OH)2}$; EG) and acetic acid ($\ce{CH3COOH}$). 

Despite the chemical complexity of Orion KL, interestingly, none of the large interstellar aldehydes, including acetaldehyde (\ce{CH3CHO}; AA) has been firmly identified at high-angular resolution. There have been several claims of acetaldehyde transitions detected toward this region~\citep{blake87, turner91, ziurys93, ikeda01, feng15}. However, in most cases, the reported spectra suffered from low S/N ratios and/or a lack of accurate spectroscopic information. Only single-dish observations by \citet{ikeda01} and \citet{charnley04} reported the clear detection of acetaldehyde transition lines so far. However, the large observing beam size ($\gtrsim16\arcsec$ or $\gtrsim$ 6200 au) and the very low $E_\textrm{u}$ of the detected lines ($\le$16 K; other bright transitions with higher $E_\textrm{u}$ were not detected) suggest that this emission might also be associated with a large-scale foreground layer of Orion KL. High-angular-resolution observations are desired to accurately map the distribution of molecular species, determine their abundances, and ultimately analyze spatial variations in the chemistry.

Acetaldehyde is of potential prebiotic importance as a possible precursor for several carbohydrates, and for larger intermediates that can form various amino acids. Recent measurements of stable $^{13}$C isotopic patterns in amino acids in early solar system bodies revealed a trend in which the $^{13}$C enrichment of $\alpha$-amino acids decreases as the carbon chain length increases. This trend has been interpreted as evidence for Strecker synthesis, in which $^{13}$C-enriched aldehyde, likely originating from the ISM or dense cloud regions, becomes diluted~\citep{chimiak21, zeichner23}. However, our understanding of acetaldehyde formation has been limited despite its presence in a range of sources. Since the first detection of acetaldehyde toward Sgr B2~\citep{gottlieb73, fourikis74}, it has been identified in various star-forming regions: from cold ($\thicksim$10 K) environments such as prestellar cores and outer envelopes of protostars~\citep{bacmann12, vastel14, jaber14, scibelliandshirley20} to warm regions, including inner protostellar envelopes and outflows~\citep{belloche13, cazaux03, imai16, bianchi19, desimone20}. Most of these detections were made by single-dish observation, and the rotational temperature estimated from these studies was quite low ($<$ 40 K) except for Sgr B2~\citep[60 K-100 K;][]{belloche13}. Recently, increasing numbers of interferometer observations with Atacama Large Millimeter/submillimeter Array (ALMA) have revealed a number of sources that host compact and warmer ($>$ 100 K) emission of acetaldehyde~\citep{lykke17, csengeri19, lee19, yang21, baek22, lee23}, and a wider range of abundance distribution is found for this species compared to other COMs~\citep{yang21}. This implies that the formation route of acetaldehyde may be sensitive to environments, and tracing the gas reservoir of the species on a smaller linear scale is critical to clarify the underlying chemistry of acetaldehyde.

In this study, we investigate an extensive ALMA archive dataset to search for acetaldehyde and other undetected aldehydes in Orion KL. Chemical modeling is also employed in this work to better understand aldehyde chemistry. In $\S$ 2, we briefly provide a summary of the investigated archive data sets. The data reduction process and chemical modeling are also described. The observational results and the derivation of the physical parameters are presented in $\S$ 3. The results of the model and discussion are provided in $\S$ 4 and $\S$ 5, respectively.

\section{Methods} \label{sec:methods}
\subsection{Observations and data reduction} \label{subsec:methods-obs}

\begin{deluxetable}{ccccc}
\tablewidth{0pt}
\tabletypesize{\footnotesize}
\tablecolumns{5}
\tablecaption{Comparison of the observational setup \label{obs_spec}}
\scriptsize
\tablehead{
\colhead{ALMA Band} & \colhead{Angular resolution} & \colhead{sensitivity\tablenotemark{a}} & \colhead{Maximum Recoverable Scale} & \colhead{The number of 12m antenna}\\
\colhead{} & \colhead{(\arcsec)} & \colhead{(mJy)} & \colhead{($\arcsec$)} & \colhead{}
}
\startdata
6 (SV) & 1.2 & 1.7-2.4 & 8.9 & 16\\
4 & 0.5 & 1.0 & 4.5 & 49\\
7 & 0.2 & 0.5 & 3.6 & 41 \\
\enddata
\tablecomments{The observational setup is based on archive data from ALMA projects 2011.0.00009.SV (Band 6), 2017.1.01149.S (Band 4), and 2016.1.00297.S (Band 7).}
\tablenotetext{a}{Estimated noise in a nominal 10 km/s bandwidth}
\end{deluxetable}

Because many studies have already been conducted with the ALMA scientific verification (SV) data of Orion KL, we analyzed four new ALMA archive datasets with an improved observational setup. They cover Band 4 (2017.1.01149) and Band 7 (2013.1.01034, 2016.1.00297, and 2016.1.01019), spanning afrequency range of 142 -- 355 GHz with a spectral resolution of 488.28~\mj{k}Hz ($\thicksim0.5\textrm{km}/\textrm{s}$). Table~\ref{obs_spec} compares the observational parameters of the ALMA archive data to those of SV data. The data were calibrated and reduced following standard routines using the Common Astronomy Software Applications (CASA) software version 6.6.3~\citep{mcmullin07}. The continuum emission was subtracted by fitting low order polynomials to carefully selected line-free channels in the image cube. For Band 7, only data obtained with a UV range greater than 35~m was considered to align with the smaller maximum recoverable scale (MRS) of Band 4. The line maps were cleaned using an auto-masking feature~\citep{kepley20} based on the {\tt hogbom} algorithm~\citep{hogbomandbrouw74} and an experimental weighting scheme that adds an inverse uv taper to Briggs weighting. The cleaned image cube was convolved with the largest synthesized beam ($0.68\arcsec\times0.50\arcsec$) of the data to achieve the consistent resolution. 

\section{Data analysis and results} \label{sec:results}
\subsection{Emission map} \label{subsec:results-emit_map}
To identify acetaldehyde emission, we select bright transitions by using the Splatalogue web tool \footnote{\url{http://www.cv.nrao.edu/php/splat}} with the following line selection criteria: Einstein coefficient log($A_\textrm{ij}$) $>$ -5 and upper level energy $E_\textrm{up} <$ 350 K. CDMS and JPL databases are used for this selection process. A total of 34 bright acetaldehyde transitions fulfilling this criterion were found within the frequency coverage. For each bright transition line, a cleaned channel map is generated for 80 channels of 0.5 km/s starting at 0~km/s defined in the LSRK frame. One of the brightest transitions of acetaldehyde at 312.78377 GHz ($16_{1,15}-15_{1,14}$, $A$; $E_\textrm{up}=131~\textrm{K}$, $S_\textrm{ij}\mu^{2}=201~\textrm{D}^{2}$) was found to be free of contamination; no bright transitions of any known interstellar species were found near ($\pm$9 MHz) the rest frequency of the transition after a search with Splatalogue. This transition is used to manually identify acetaldehyde components, after which a deeper investigation was performed for the identified regions.

\begin{figure*}
\gridline{\fig{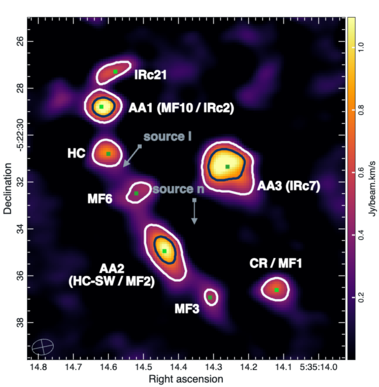}{0.5\textwidth}{}
          \fig{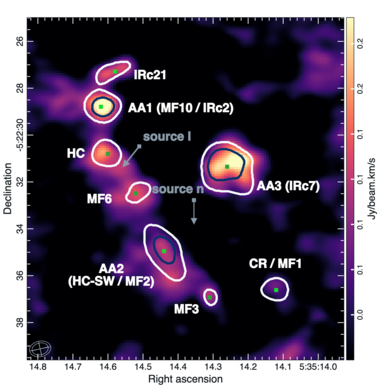}{0.5\textwidth}{}}
   \caption{Acetaldehyde integrated intensity maps at Band 7 (left) and Band 4 (right). The synthesized beam is shown in the bottom left corner. The contour levels are (5,10) $\times$ sigma ($\sigma$=0.07 Jy/beam~km/s) and the contour derived from the Band 7 data is overlaid on the Band 4 map for comparison. Green dots mark the AA peak positions determined from a 2-D Gaussian fitting of the map in the left panel. Velocity vectors representing the proper motion of source I and n are shown as gray arrows.}
\label{figure-mom0_map}
\end{figure*}

\begin{figure}
\centering
\includegraphics[width=0.7\textwidth]{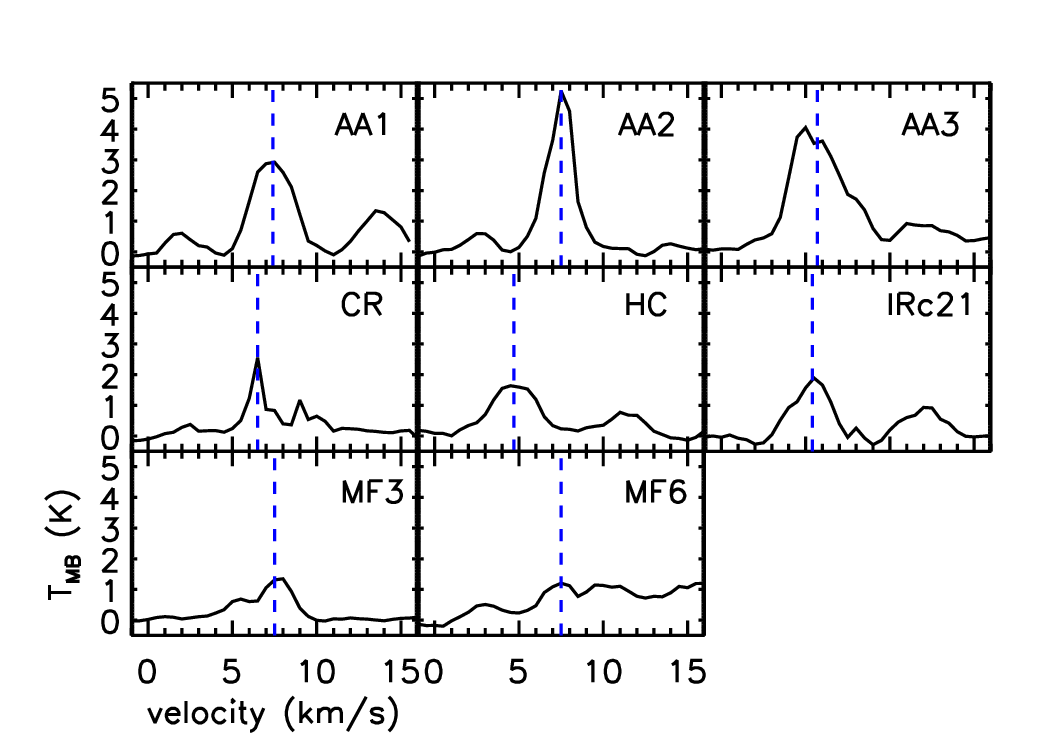}
\caption{The spectra of the representative line (312.78377 GHz) at the eight acetaldehyde peaks identified above the 5$\sigma$ threshold.}
\label{figure-312.78377ghz_all_rgns}
\end{figure}
\begin{figure*}
\centering
\begin{minipage}{0.45\textwidth}
  \centering
  \includegraphics[width=87mm]{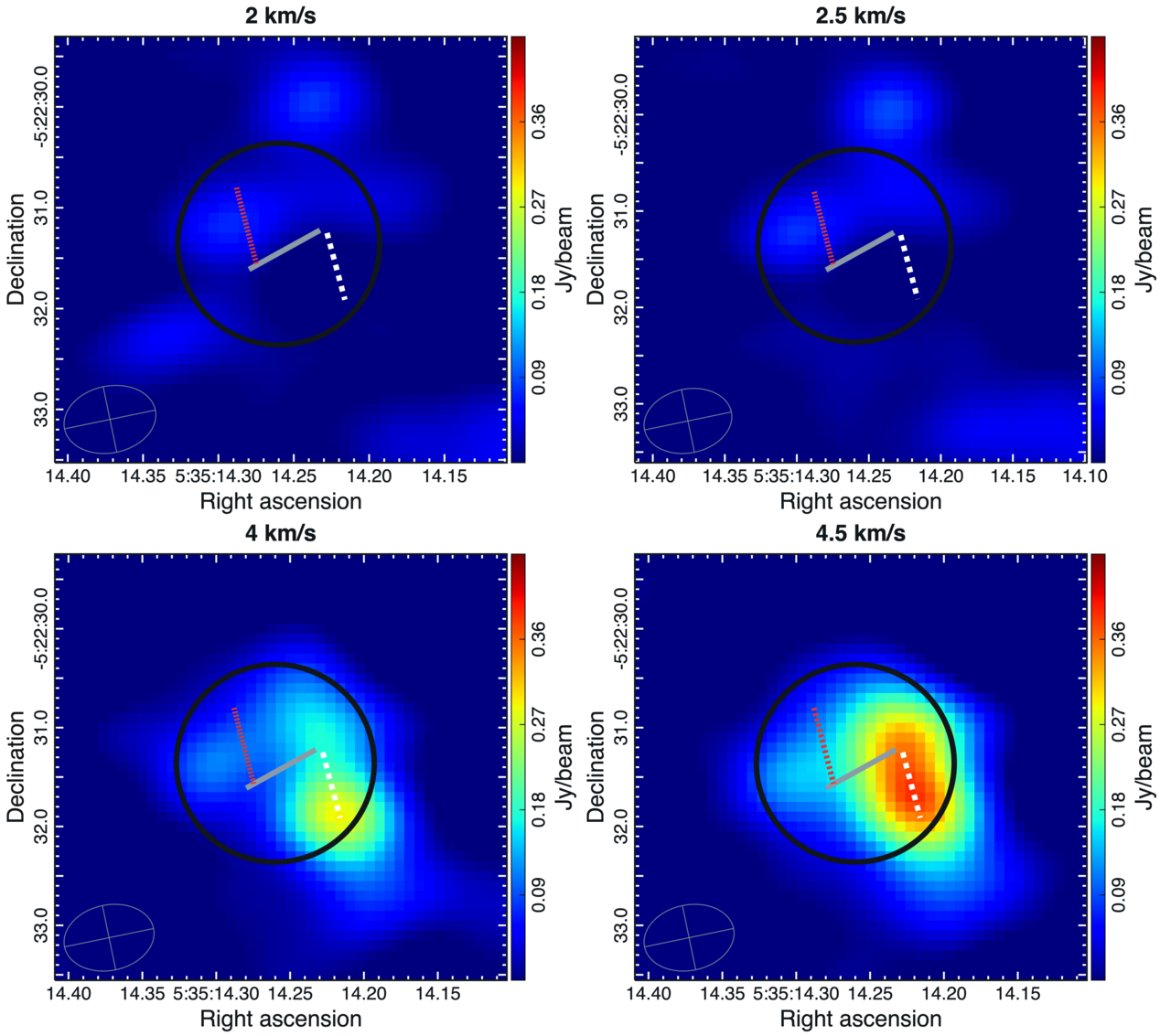}
\end{minipage}
\hspace{0.00\textwidth} 
\begin{minipage}{0.45\textwidth}
  \centering
  \includegraphics[width=87mm]{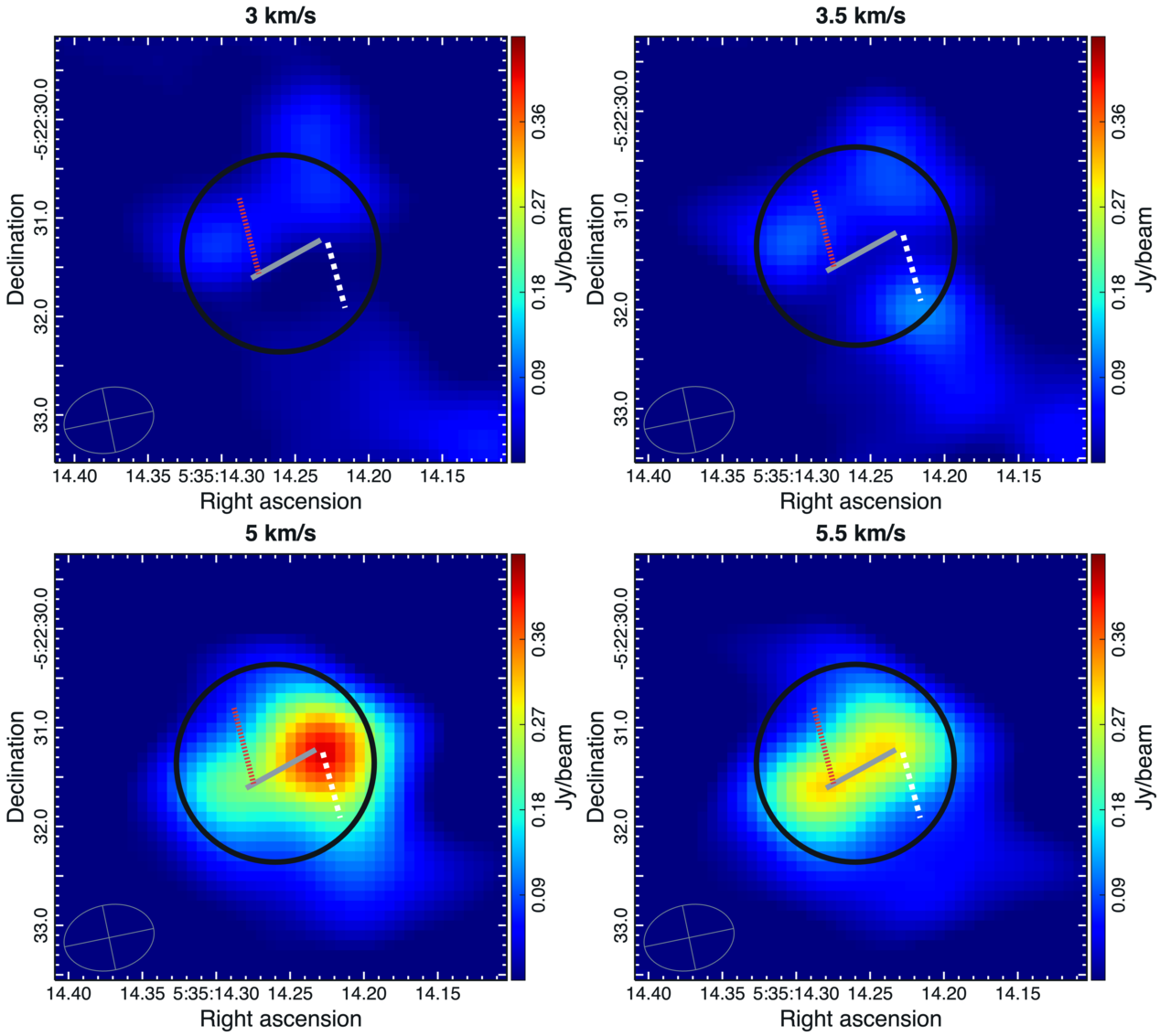}
\end{minipage}

\vspace{0.0\textwidth} 

\begin{minipage}{0.45\textwidth}
  \centering
  \includegraphics[width=87mm]{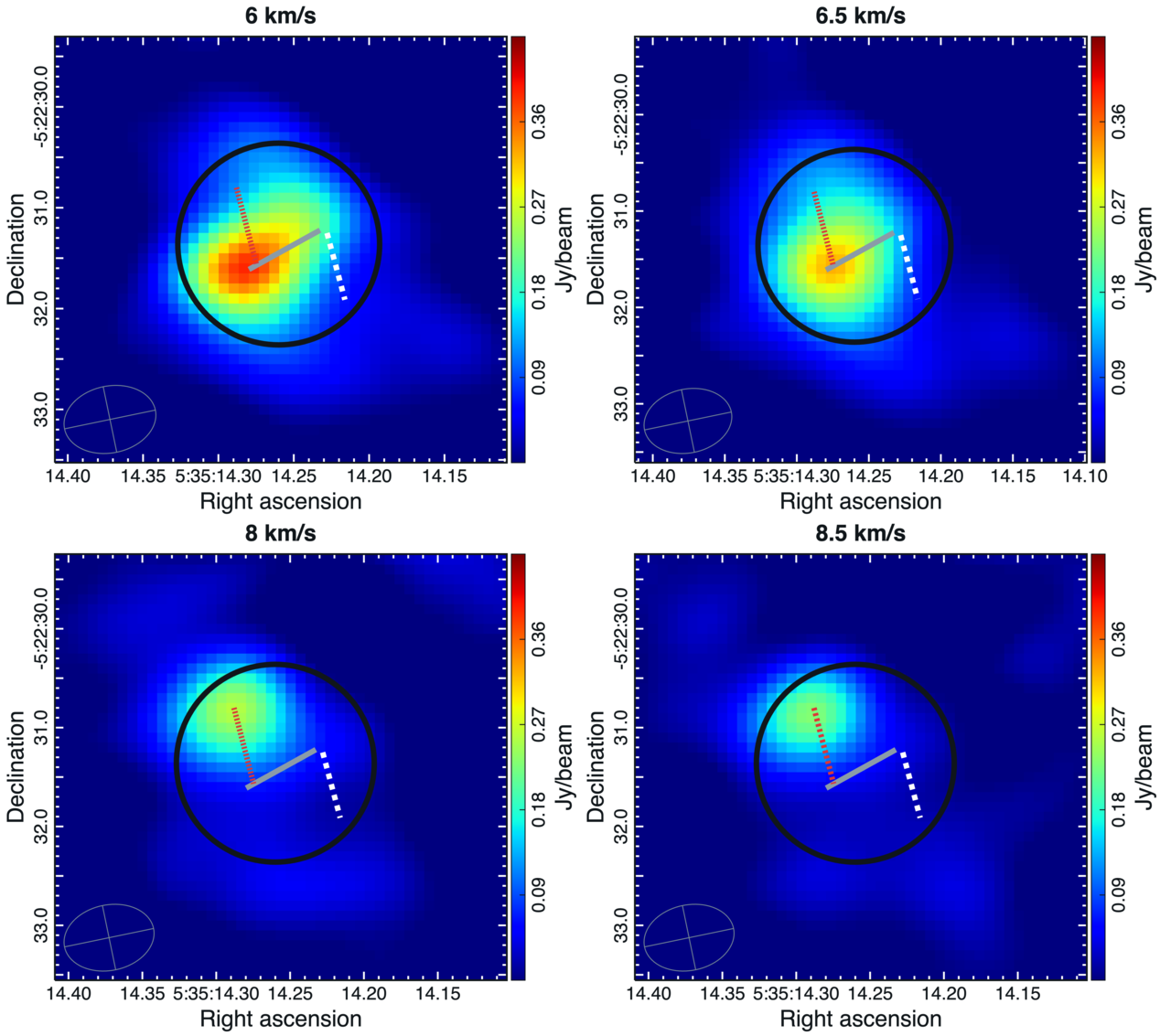}
\end{minipage}
\hspace{0.00\textwidth}
\begin{minipage}{0.45\textwidth}
  \centering
  \includegraphics[width=87mm]{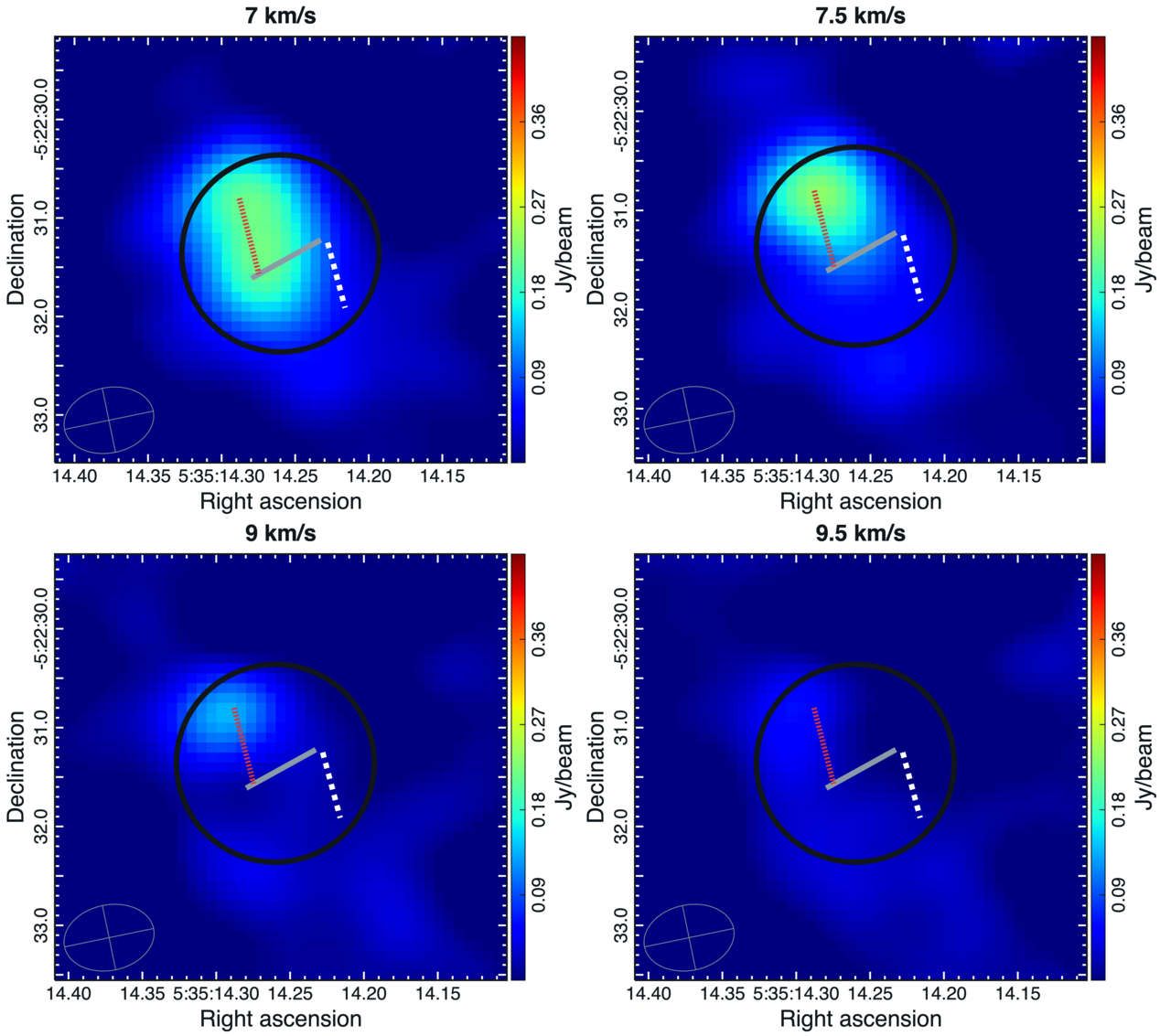}
\end{minipage}

\caption{Channel map of 312.78377 GHz representative line toward AA3. The black circle indicates the region used for spectral line extraction. The white-dashed, gray-solid, and red-dotted lines represent the velocity gradient axes for the 4.5, 5.5, and 7 km/s $v_\textrm{LSR}$ components in AA3, respectively.}
\label{figure-chan_map}
\end{figure*}

As seen in the left panel of Figure~\ref{figure-mom0_map}, moment 0 maps are generated by integrating the velocity channels of both this representative acetaldehyde transition and two blended lower-energy transitions (154.32220 GHz; $E_\textrm{u}=53 K$) from $v_\textrm{LSR}=0.0$ km/s to $v_\textrm{LSR}=10.5$ km/s, ensuring that they encompass various velocity components. The emission distributions of these two transitions are highly similar, confirming that the detected emission components are real. Additionally, acetaldehyde emission from Cycle 0 Band 6 data with shorter baselines also exhibits a similar distribution~(Liu, priv. comm.). Three emission peaks (AA1 to AA3; dark-blue contour) were identified using a 10 $\sigma$ threshold, and five peaks above 5 $\sigma$ (white contour) are also shown in the figure. All of them are associated with previously reported sources -- MF10/IRc2 (AA1), HC-SW - Ethanol Peak (AA2), IRc7 (AA3), CR/MF1, HC, IRc21/HC-N, MF6, and MF3. Figure~\ref{figure-312.78377ghz_all_rgns} presents the spectrum of the representative acetaldehyde transition extracted from the eight distinct regions. The brightest peak, AA1, aligns with MF10/IRc2. Notably, while emission from other O-bearing COMs in this region has been sparsely documented, strong features of N-bearing COMs, such as \ce{CH2CHCN} and \ce{CH3CH2CN}, have been previously reported~\citep[e.g.][]{cernicharo16, pagani17, margules18}. This position exhibits two velocity components centered at 7.5 km/s and 1.9 km/s~\citep{pagani17} The second brightest peak, AA2, is located approximately 4$\arcsec$.5 southwest of the hot core. This HC-SW region is known for hosting the most abundant chemical inventory of organic materials in Orion KL, including many COM sources like the ethylene glycol peak and the ethanol peak (EtP)~\citep{tercero15, tercero18}. In particular, AA2 corresponds to the ethanol peak, suggesting a chemical correlation between acetaldehyde and ethanol and/or a similarity in their excitation behaviors. Although not as chemically rich as AA2, AA3 still exhibits emissions from several N-and O-bearing COMs~\citep{cernicharo16, pagani17, margules18, tercero18}. This region has a complex kinematic structure. Figure~\ref{figure-chan_map} presents the acetaldehyde emission map as a function of velocity, showing a peak that moves in a zigzag pattern (4-7.5km/s). This pattern is likely caused by the geometric projection of a single large kinematic structure, such as a bipolar outflow. Since this variation is smaller than the beam size used for spectral extraction, we did not conducted further spectral analysis on this position.

\begin{deluxetable}{cccccc}
\tablewidth{0pt}
\tabletypesize{\footnotesize}
\tablecolumns{6}
\tablecaption{Source parameters characterized by 2-D Gaussian map ftting \label{table-2-d_fit}}
\scriptsize
\tablehead{
\colhead{AA component} & \colhead{position name} & \colhead{RA (J2000)} &  \colhead{Dec (J2000)} &  \colhead{$V_\textrm{LSR}$} &  \colhead{source size} \\
\colhead{} & \colhead{} & \colhead{$05^\textrm{h}35^\textrm{m}$...} &  \colhead{-$05^{\arcdeg}22^{\arcmin}$...} &  \colhead{(km/s)} &  \colhead{($\arcsec\times\arcsec$)}  
}
\startdata
AA1 & MF10/IRc2	&	14.62	&	28.79	&	1.9, 7.5	&	1.2$\times$1.0	\\
AA2 & HC-SW (Ethanol Peak; EtP)	&	14.44	&	34.95	&	7.5	&	2.2$\times$1.1	\\
AA3 & HC-W(the vicinity of IRc7)	&	14.26	&	31.35	&	4.5-6.5	&	1.9$\times$1.7	\\
 & CR (MF1)	&	14.12	&	36.61	&	6.5	&	1.3$\times$1.1	\\
 & HC	&	14.60	&	30.80	&	4.5	&	1.3$\times$1.1	\\
 & IRc21 (HC-N)	&	14.58	&	27.29	&	6	&	1.8$\times$0.7	\\
 & MF3      &   14.31       &   36.93       &   7.5 &   1.7$\times$1.0 \\
 & MF6      &   14.52  &     32.49&   7.5 &   1.7$\times$1.6 \\ 
\enddata
\end{deluxetable}

Acetaldehyde source parameters are estimated using a 2-D Gaussian fitting method, and the results is summarized in Table~\ref{table-2-d_fit}. It should be noted that the substructures within AA3 are not resolved in this analysis. We fit AA3 as a single component by integrating over a relatively wide velocity interval ($v_\textrm{LSR}$ from 0.0~km/s to 10.5 km/s). Separately fitting the substructures with a narrower velocity integration interval was disfavored due to the complex spatial and kinematic structure of AA3. Given these complexities, our subsequent analysis primarily focuses on the other two sources.

\subsection{Line Identification and Spectra} \label{subsec:results-spectra}
Other acetaldehyde transitions are investigated by analyzing their spectra extracted from the two intensity peaks of acetaldehyde. In the spectra, the line peaks are compared with the rms noise levels ($\sigma$) to determine line detection significance. We performed a 1-D Gaussian fitting for the collected spectra using the \textit{specfit} task in CASA. The rms was measured using CASA \textit{imstat} across the region excluding emission in a specific channel where the line peak exists. This line detection analysis was performed without correction for primary beam attenuation to achieve consistent noise levels across the field of view. 

\begin{deluxetable}{ccccccccccc}
\tablewidth{0pt}
\tabletypesize{\footnotesize}
\tablecolumns{11}
\tablecaption{Line parameters for the detected transitions for AA2 \label{table-spec_fit}}
\scriptsize
\tablehead{
\colhead{Band} & \colhead{Frequency} & \colhead{Transition} & \colhead{$S_\textrm{ij}\mu^{2}$} & \colhead{log($A_\textrm{ij}$)} &  \colhead{$E_\textrm{up}$} & \colhead{$T_\textrm{peak}$}  & \colhead{$\Delta v$} & \colhead{Integral} & \colhead{$T/S_{\nu}$} & \colhead{comment} \\
\colhead{} & \colhead{(GHz)} & \colhead{} & \colhead{($D^{2}$)} & \colhead{(s$^{-1}$)} & \colhead{(K)} &\colhead{(mJy/beam)}  & \colhead{(km/s)} & \colhead{(mJy/beam km/s)} & \colhead{(K beam /Jy)} & \colhead{}
}
\startdata
4 & 154.32220\tablenotemark{a} & 8(-3, 6)- 7(-3, 5) E & 87 & -4.0 & 54 & 79.2 (5.8)& 2.2 (0.2) & 185.9 (13.5) & 147 & a-type \\
7 & 312.71061 & 16( 1,15)-15( 1,14) E & 201 & -3.0 & 131 & 509.9 (40.1) & 1.8 (0.12) & 954.0 (75.0) & 20 & a-type \\
7 & 312.78377 & 16( 1,15)-15( 1,14) A  & 201 & -3.0 & 131 & 529.2 (173.7) & 1.8 (0.7) & 1019.4 (334.7) & 20 & a-type \\
7 & 354.81294 & 18( 2,16)-17( 2,15) E & 225 & -2.8 & 170 & 791.9 (132.6) & 1.1 (0.2) & 956.8 (160.0) & 18 & a-type \\
\enddata
\tablenotetext{a}{This line is blended with another acetaldehyde transition at 154.32254~GHz, which has a comparable upper-state energy ($E_\textrm{up}$) and line strength (8(3,5)-7(3,4) A).}
\end{deluxetable}

Four acetaldehyde transitions, including the representative line at 312.78377 GHz, are detected above the 5 $\sigma$ level out of 34 acetaldehyde transitions that meet the criteria for expected bright molecular emission. All detected lines in Band 7 are free from severe blending, while the Band 4 line is blended with another acetaldehyde transition of similar $E_\textrm{up}$ and line strength (154.32254~GHz or $\Delta v\sim-0.7$ km/s; 8(3,5)-7(3,4) A). This blending results in a broader linewidth compared to the other transitions. Table~\ref{table-spec_fit} summarizes the line parameters and associated $\sigma$ values determined from the spectra collected from the intensity peak pixel of AA2. The detected transitions are found in the two project datasets -- 2016.1.00297 (Band 7) and 2017.1.01149 (Band 4) out of the four projects analyzed. All detected lines are associated with intrinsically brighter transitions compared to the non-detected ones. The detected lines are characterized by $S_\textrm{ij}\mu^{2}\thicksim90~\textrm{D}^{2}$ and $E_\textrm{up}\thicksim~60~\textrm{K}$ in Band 4, $S_\textrm{ij}\mu^{2}>200~\textrm{D}^{2}$ and $E_\textrm{up}< 170~\textrm{K}$ in Band 7. LTE line modeling indicates that seven additional transitions should be detectable based on their line flux; however, all of these transitions exhibit severe non-Gaussian profiles, indicating contamination from nearby transitions or complex velocity structures~(see Figure~\ref{figure-a1}). Of the 34 bright transitions, the four detected lines listed, along with the seven transitions predicted to be detectable but excluded due to blending, are all a-type transitions ($\Delta\kappa_\textrm{p}=0$, $\Delta\kappa_\textrm{o}=\pm1$). The remaining undetected transitions belong to other types, emphasizing the brightness advantage of a-type transitions for this species.

Three out of four transitions are encompassed in Band 7, while a single transition is found in Band 4, covering the low-energy range ($E_\textrm{up}=53$ K). The lines show similar spatial morphology despite having different upper-state energy levels. The moment 0 map of this low-energy transition is presented in the right panel of Figure~\ref{figure-mom0_map}.

\subsection{Derivation of Physical Parameters} \label{subsec:results-spec_model}
\begin{figure}
\centering
\includegraphics[width=\textwidth]{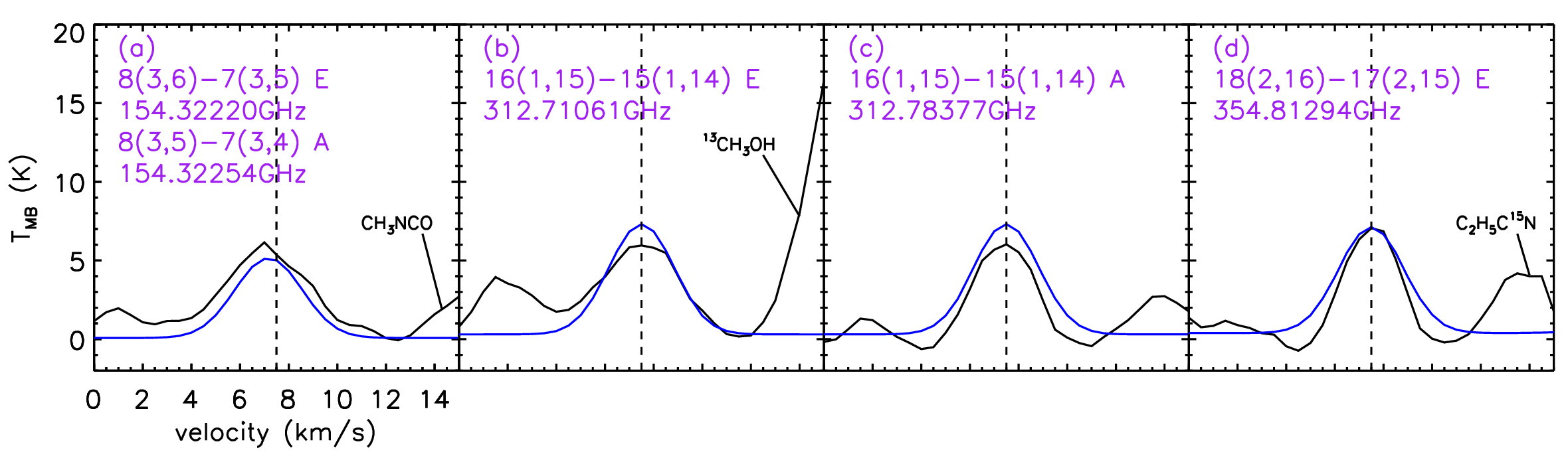}
\caption{The XCLASS model spectra of acetaldehyde (blue) overlaid with ALMA observations toward AA1 (black solid).}
\label{figure-spec_fit_aa1}
\end{figure}

\begin{figure}
\centering
\includegraphics[width=\textwidth]{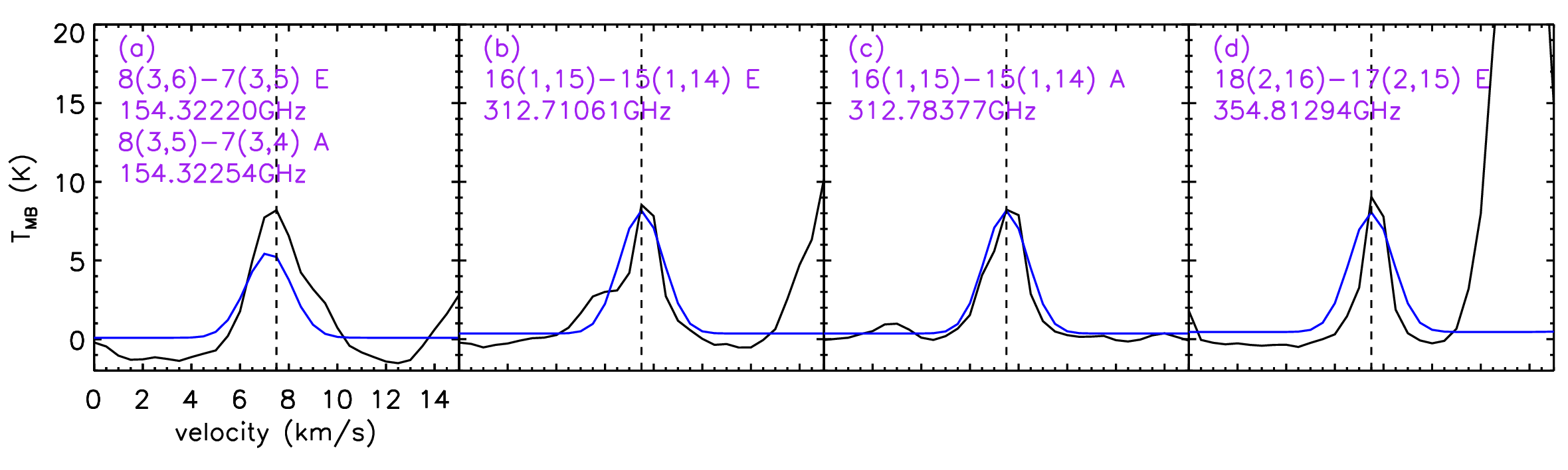}
\caption{The XCLASS model spectra of acetaldehyde (blue) overlaid with ALMA observations toward AA2 (black solid) }
\label{figure-spec_fit_aa2}
\end{figure}

Molecular lines from the two main emission regions, AA1 and AA2, are modeled using the eXtended CASA Line Analysis Software Suite (XCLASS)~\citep{moller17}. The observations are compared with the synthesized spectra to find the column densities~($N_\textrm{tot}$) and the excitation temperatures~($T_\textrm{rot}$) that best fit the observation. Figure~\ref{figure-spec_fit_aa1} and \ref{figure-spec_fit_aa2} show the best-fit results of the AA lines. It takes into account beam dilution and line opacity under the assumption of local thermodynamic equilibrium (LTE). The column densities and rotational temperatures of acetaldehyde estimated toward AA2 will be discussed in \S 4 by comparison with those of other HCO-bearing species from TCL18.

The main input parameters for the spectral modeling are systemic velocity $v_\textrm{LSR}$, source size, line FWHM $\Delta v$, rotational temperature $T_\textrm{rot}$, and column density $N_\textrm{tot}$. We assume $N(\textrm{H})=2.0\times10^{24}~\textrm{cm}^{-2}$ and $v_\textrm{LSR}$=7.5 km/s~\citep{pagani17}. The source sizes of AA1 and AA2 are assumed to be  1.1$\arcsec$ and 1.5$\arcsec$ respectively, which are equivalent to the effective source sizes  based on the 2-D Gaussian fitting results. The other three parameters are left free. The spectrum of acetaldehyde is first computed with an initial guess of parameter vectors. These are the line width ($\Delta v$) of 2-3 km/s, $T_\textrm{ex}$ of 80-100 K, and $N_\textrm{tot}$ of $2.0\times10^{15}~\textrm{cm}^{-2}$, depending on the source. The model then finds the chi-squared minimum by varying $\Delta v$ between 1.0 km/s and 3.5 km/s, column density between $10^{14}~\textrm{cm}^{-2}$ and $10^{17}~\textrm{cm}^{-2}$, and excitation temperature between 50 K and 150 K, with estimated fitting errors. The parameter space is explored through a combination of three optimization algorithms; these are genetic, Levenberg-Marquardt to find the best solution, and the Markov chain Monte Carlo (MCMC) method for the error estimation~\citep[see][and references therein for more detail]{moller17}.

To provide appropriate spectra to be compared with the model, a primary beam correction is performed for the acetaldehyde-detected channel maps. Spectra are then re-extracted by averaging over a $1\arcsec\times1\arcsec$ aperture around the two emission peaks, AA1 and AA2. This extraction region size is chosen to be consistent with TCL18 to facilitate comparison with other COMs at AA2. The flux density is converted to an antenna temperature using the conversion factor appropriate to its spectral window as listed in Table~\ref{table-spec_fit}. The best-fit results are summarized in Table~\ref{table-xclass_fit}.

\begin{deluxetable}{cccc}
\tablewidth{0pt}
\tabletypesize{\footnotesize}
\tablecolumns{4}
\tablecaption{XCLASS fitting results of AA1 and AA2 \label{table-xclass_fit}}
\scriptsize
\tablehead{
\colhead{AA component} & \colhead{position name} &  \colhead{$N_\textrm{tot}$(\ce{CH3CHO})} & \colhead{$T_\textrm{ex}$} \\
\colhead{} & \colhead{} & \colhead{($10^{15}~$cm$^{-2}$)} & \colhead{(K)}   
}
\startdata
AA1 & MF10/IRc2	&	$3.1_{-0.46}^{+1.41}$	 & $107_{-23}^{+3}$ \\
AA2 & HC-SW (Ethanol Peak; EtP)	&	$2.9_{-1.2}^{+0.4}$  & $142_{-1}^{+3}$ \\
\enddata
\end{deluxetable}

The modeled temperature in AA1 is lower than in AA2 while the column densities are similar. However, it is important to note that the temperature toward AA1 should be considered carefully. The derived temperatures are sensitive to the initial guess and the range of temperatures explored, whereas the column densities remain robust. This uncertainty might be due to potential non-LTE behavior or the presence of multiple components with different temperatures when explosive shocks or external heating excite acetaldehyde transitions in a low-density environment.

\subsection{Column density ratios at AA2} \label{subsec:results-chemaa2}

\begin{deluxetable}{ccc}
\tablewidth{0pt}
\tabletypesize{\footnotesize}
\tablecolumns{3}
\tablecaption{Column densities of HCO-bearing species \label{table-coldens}}
\scriptsize
\tablehead{
\colhead{Species} & \colhead{$N_\textrm{tot}\textrm{(i)}$} & \colhead{$\frac{N_\textrm{tot}\textrm{(i)}}{N_\textrm{tot}\textrm{(FA)}}$}\\
\colhead{} & \colhead{($\textrm{cm}^{-2}$)} 
}
\startdata
\ce{HCOOH}; FA & 1.0E+15 & 1\\
\ce{CH3CHO}; AA & 3.0E+15 & 3\\
\ce{HCOOCH3}; MF & 2.4E+17 & 240\\
\ce{CH2OHCHO}; GA & $<$ 3.0E+14 & $<$0.3\\
\ce{C2H5OH}; EtOH & 6.4E+16 & 64 \\
\enddata
\end{deluxetable}

The relative abundance ratios between the HCO-bearing species and ethanol at AA2 are determined based on the results of this work and TCL18. TCL18 investigated the chemical composition of thirteen O-bearing COMs in HC-SW and CR including ethanol and three species with an -HCO moiety (\ce{HCOOCH3} methyl formate; MF, \ce{CH2OHCHO} glycolaldehyde; GA, \ce{HCOOH} formic acid; FA, and \ce{HCOOC2H5} ethyl formate; EF) using ALMA SV data. Table~\ref{table-coldens} lists the column densities for the species of interest estimated from our work and TCL18. The relative abundance ratios are derived by normalizing with the column density of formic acid, as it is the smallest reliably measured value among the five species of interest, aside from the upper limit set for glycolaldehyde.

The column densities observed are significantly different depending on the species, showing a wide range of relative abundance ratios from $<$ 0.3 to 240. The angular resolution and MRS of the archive data investigated here are about two times smaller than those of the SV data used by TCL18. This means that the large-scale structure previously seen in the TCL18~(SV data) can be resolved out in our study. However, to our knowledge, only AA transitions whose upper energy is less than 20~K have been clearly detected toward Orion KL with single-dish observation. Other bright transitions with higher $E_{u}$ were not detected in \citet{ikeda01}. This implies that the flux contribution of the large-scale acetaldehyde structure for the transitions of our interest ($E_\textrm{u}~>~50~$K) is minor and the effect of filtering out would be negligible. In addition, the region size for the spectral extraction is smaller than the MRS.

\section{Aldehyde chemistry and chemical modeling} \label{sec:discuss-chem_model}

\subsection{Chemical model} \label{subsec:methods-model}

We perform a one-point chemical model for HC-SW that is associated with the bright acetaldehyde emission peak, AA2. The observed relative abundance ratios are investigated using the astrochemical kinetic code {\em MAGICKAL} to understand the chemistry related to the aldehyde-like species and ethanol. This code solves a set of rate equations describing a fully coupled gas-phase, grain-surface, and ice-mantle chemistry under time-dependent physical conditions appropriate to the source~\citep{garrodandpauly11, garrod13}. The details of this version of {\em MAGICKAL} are described in Appendix~\ref{appendix:magickal}.

The chemical evolution is initiated from the typical elemental and chemical abundances of dark clouds~(Table~\ref{init-element}). The physical evolution in this model is described with a two-stage approach, in which (i) an initial, cold core is formed and then (ii) it is warmed up to a high temperature. Given the complex environment of Orion KL, it is possible that other energetic events occurred before the recent explosion, leading to chemistry that was more evolved and complex than the model assumed for the pre-explosion environment. However, as considering this variable with the lack of information is beyond the scope of this work, here we focus on the simplest case, and the impact of possibly more evolved chemistry before the recent explosive event will be discussed in \S~\ref{subsubsec:aa_chem}.

During the first stage, the gas density ($n_\textrm{H}$) rises from a value of 3000~cm$^{-3}$ to the value measured toward the main emission region in Orion KL ~($n_\textrm{H}\thicksim8\times10^{8}~\textrm{cm}^{-3}$). This density evolution basically follows the free-fall collapse timescale, but it can slow down by employing a magnetic field retardation factor, $B$. The dust temperature $T_\textrm{d}$ during the collapse stage is calculated according to the visual extinction-dependent formula provided by \citet{garrodandpauly11}. The gas temperature is held steady at 10~K throughout the collapse stage. The warm-up stage then follows with a timescale of 550 years based on the assumption that this region is externally heated by the recent explosion. Here, $T_\textrm{kin}$ increases from 8 K to a range of maximum temperatures (100~K - 150~K) to understand the effect of temperature on chemistry. The assumed physical conditions are based on \citet{pagani17} and references therein.

\begin{deluxetable}{cc}
\tablewidth{0pt}
\tabletypesize{\footnotesize}
\tablecolumns{2}
\tablecaption{Initial elemental and chemical abundances \label{init-element}}
\scriptsize
\tablehead{
\colhead{Species, \textit{i}} & \colhead{$n(i)/n_\textrm{H}$ \tablenotemark{a}
}}
\startdata
$\ce{H}$ & 5.0(-4) \\
$\ce{H2}$ & 0.49975\\
$\ce{He}$ & 0.09\\
$\ce{C}$ & 1.4(-4)\\ 
$\ce{N}$ & 2.1(-5)\\
$\ce{O}$ & 3.2(-4)\\
$\ce{S}$ & 8.0(-8)\\
$\ce{Na}$ & 2.0(-8)\\
$\ce{Mg}$ & 7.0(-9)\\
$\ce{Si}$ & 8.0(-9)\\
$\ce{P}$ & 3.0(-9)\\
$\ce{Cl}$ & 4.0(-9)\\
$\ce{Fe}$ & 3.0(-9)\\
\enddata
\tablecomments{$^{a}A(B)=A^{B}$}
\end{deluxetable}

The visual extinction during the collapse takes an initial value of 2 mag, and is scaled with the density according to the expression $A_\textrm{V} \propto n_{H}^{2/3}$. An additional background visual extinction ($A_\textrm{V,BAC}$) can be considered under the assumption that the collapse is initiated in the deeply embedded position in the molecular cloud. Once the density reaches the final value specified for the source under consideration, it is followed by a static warm-up stage, which resembles the vicinity of a heat source such as an explosion. The warm-up characteristics, such as maximum temperature and warm-up timescale, can be tuned depending on the source properties. The gas density is assumed to be high enough for the gas and dust temperatures to be well coupled at this stage.

\begin{deluxetable}{ccc}
\tablewidth{0pt}
\tabletypesize{\footnotesize}
\tablecolumns{3}
\tablecaption{Ice abundances of the {\tt control} model at the end of the collapse stage \label{table-control_ice_abun}}
\scriptsize
\tablehead{
\colhead{Species} & \colhead{$n(i)/n_\textrm{H}$} & \colhead{$\frac{n\textrm{(i)}}{n\textrm{(FA)}}$}
}
\startdata
FA & 1.4E-07 & 1 \\
AA & 5.5E-09 & $<$0.1 \\
MF & 1.2E-07 & 0.9 \\
GA & 3.0E-08 & 0.2 \\
EtOH & 3.3E-07 & 2.4 \\
\enddata
\end{deluxetable}

\begin{deluxetable}{ccc}
\tablewidth{0pt}
\tabletypesize{\footnotesize}
\tablecolumns{3}
\tablecaption{Gas-phase abundances of the {\tt control} model at the end of warm-up ($T_\textrm{kin,max}$=120 K)\label{table-control_gas_abun}}
\scriptsize
\tablehead{
\colhead{Species} & \colhead{$n(i)/n_\textrm{H}$} & \colhead{$\frac{n\textrm{(i)}}{n\textrm{(FA)}}$}
}
\startdata
FA & 9.3E-10 & 1 \\
AA & 5.6E-11 & 0.1 \\
MF & 1.3E-08 & 14.0 \\
GA & 1.9E-12 & $<$0.1 \\
EtOH & 2.2E-11 & $<$0.1 \\
\enddata
\end{deluxetable}

\begin{figure}
\centering
\includegraphics[width=0.7\textwidth]{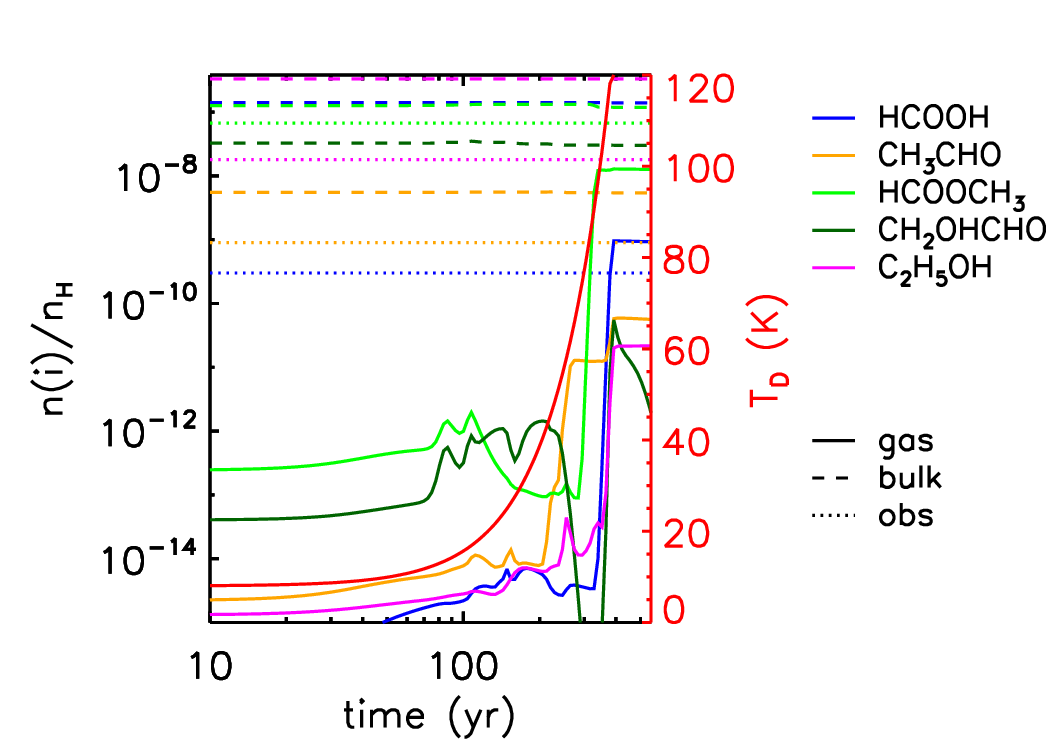}
\caption{Time evolution of abundances of COMs in the warm-up phase of the {\tt control} model. Abundances in the gas and in the bulk ice are denoted by solid and dashed lines, respectively. The gas-phase abundances from \citet{tercero18} are shown by dotted lines. The upper limit of the gas-phase abundance of glycolaldehyde (\ce{CH2OHCHO}) is $\le~4\times10^{-10}$ and is not presented in this figure.}
\label{figure-warmup_ctrl}
\end{figure}

\subsubsection{{\tt control} model}

We first set up a {\tt control} model and then explore how the various physical and chemical inputs of the model affect the chemistry to find the best model. The essential model setup and a chemical reaction network for the {\tt control} model are consistent with a ``final” model setup described by \citet{garrod22}. The chemical network, final binding energy and diffusion barrier treatments are adopted from this model as well. A free-fall collapse timescale (no magnetic retardation) is assumed. The collapse occurs at an initial visual extinction of $A_\textrm{V,0}$=2. Table~\ref{table-control_ice_abun} summarizes the ice abundances modeled at the end of density evolution, while Table~\ref{table-control_gas_abun} lists the gas-phase abundances of species when the temperature reaches $T_\textrm{max}$ after warm-up. The abundances normalized to the simplest species of interest, formic acid, are also presented for a quick comparison between species. Figure~\ref{figure-warmup_ctrl} shows the chemical evolution of the {\tt control} model with warm-up, and it should be noted that a significant fraction of ice species are retained in the ice. The calculated gas-phase abundances of acetaldehyde and ethanol after the warm-up are a few orders of magnitude smaller than those from observations, implying the necessity of fine-tuning the model.

In this study, we examined four different model inputs -- chemical reaction networks, visual extinction, collapse timescale, and binding energies of the species. Through this modeling work, we aim to 1) explain the observed GA:FA:AA:MF ratio in the AA2 region, 2) explore dominant formation and destruction reactions for each species, and 3) understand the chemical connection between species that exhibit spatial correlation, as discussed in~\ref{subsec:discuss-spat_distribution}.

\subsubsection{Fine tuning of the model}

\paragraph{Chemical network} \label{subsubsec:chem_network}

\begin{deluxetable}{clccc}
\tablewidth{0pt}
\tabletypesize{\footnotesize}
\tablecolumns{5}
\tablecaption{Updated / newly-included gas-phase reactions \label{table-gas_rxns}}
\scriptsize
\tablehead{
\colhead{Related species} & \colhead{Reaction} & \colhead{$k(T)$} & \colhead{$k_\textrm{OH}$ (120K)} & \colhead{Notes}\\
\colhead{} & \colhead{} & \colhead{(cm$^{3}$/s)} & \colhead{(cm$^{3}$/s)} & \colhead{}
}
\startdata
MF & $\ce{OH}+\ce{HCOOCH3} \rightarrow \ce{COOCH3} + \ce{H2O}$ & & 5.0E-13 & \tablenotemark{a}\\
FA & $\ce{OH}+\ce{HCOOH} \rightarrow \ce{COOH} + \ce{H2O}$ & & 5.0E-13 & \tablenotemark{b}\\
AA & $\ce{OH}+\ce{CH3CHO} \rightarrow \ce{CH3CO} + \ce{H2O}$ & $1.2\times10^{-11}(T/300~\textrm{K})^{-1.8}e^{-28.7/T}$ & 5.0E-11 & \tablenotemark{c} \\
EtOH & $\ce{OH}+\ce{C2H5OH} \rightarrow \ce{C2H4OH} + \ce{H2O}$ & $2.0\times10^{-11}(T/300~\textrm{K})^{-0.71}$ & 4.0E-11 &  \tablenotemark{d}  \\
\enddata
\tablenotetext{a}{\citet{jimenez16, lecalve97}; The rate coefficient is derived by interpolating the data sets measured at two temperature regime, 22-64 K and  233-372 K. For more details, see Appendix~\ref{appendix:kmf}}
\tablenotetext{b}{The $k_\textrm{MF} (120K)$ is adopted.}
\tablenotetext{c}{\citet{blazquez20}; 11.7-115 K}
\tablenotetext{d}{\citet{ocana18}; 21-107 K}
\end{deluxetable}

\begin{deluxetable}{clccc}
\tablewidth{0pt}
\tabletypesize{\footnotesize}
\tablecolumns{5}
\tablecaption{Updated / newly-included grain-surface and ice-mantle reactions \label{table-solid_rxns}}
\scriptsize
\tablehead{
\colhead{Related species} & \colhead{Reaction} & \colhead{$E_\textrm{A}$ (K)} & \colhead{Width (\AA)} & \colhead{Notes}}
\startdata
FA & $\ce{CO}+\ce{OH} \rightarrow \ce{CO2}$ + H / COOH & 800 & 1.0 & branching ratio updated -  99:1 $\rightarrow$ 99.9:0.1  \\
EtOH & $\ce{H}+\ce{C2H2} \rightarrow \ce{C2H} + \ce{H2}$ & 1300 & 1.0 & new \tablenotemark{a} \\
     & $\ce{H}+\ce{C2H3} \rightarrow \ce{C2H2} + \ce{H2}$ & 0  & 1.0 & new, 50$\%$ fraction against hydrogenation  \\%
     & $\ce{H}+\ce{C2H4} \rightarrow \ce{C2H5}$ & 1410  & 1.0 & $E_\textrm{A}$ updated - 605 K $\rightarrow$ 1410 K  \tablenotemark{b} \\
     & $\ce{O}+\ce{C2H5} \rightarrow \ce{C2H4} + \ce{OH}$ & 0 & 1.0 & new, 50$\%$ fraction against hydrogenation \\
     & $\ce{OH}+\ce{C2H5} \rightarrow \ce{H2O} + \ce{C2H4}$ & 0 & 1.0 & new, 50$\%$ fraction against hydrogenation \\
AA & $\ce{H}+\ce{CH3CO} \rightarrow \ce{CH2CO} + \ce{H2}$ & 0  & 1.0 & new, 50$\%$ fraction against hydrogenation \\
EtOH, AA & $\ce{H}+\ce{C2H4OH} \rightarrow \ce{CH3CHO} + \ce{H2}$ & 0 & 1.0 & new, 5$\%$ fraction against other reaction channels. 
\enddata
\tablenotetext{a}{The same values for H+\ce{C2H2} $\rightarrow$ \ce{C2H3} ; \citet{baulch92}}
\tablenotetext{b}{\citet{kerrandparsonage72}}
\end{deluxetable}

To explain the relative abundance ratios from the observation, we add or update the reactions related to the HCO-bearing species and ethanol based on the chemical network of \citet{garrod22}. The new/updated reactions in the gas and solid phases to this network are summarized in Table~\ref{table-gas_rxns}-\ref{table-solid_rxns}, respectively. The detailed description of this update is provided in Appendix~\ref{appendix:kmf} and \ref{appendix:solid_network}. The gas-phase reactions newly included here are the neutral-neutral reactions where the four species of interest (MF, FA, AA, and EtOH) are destroyed through H-abstraction by OH. Due to quantum tunneling, the rate coefficients of these reactions have been recently estimated and are found to be more efficient at an astrophysically relevant temperature than was previously expected~\citep{ocana18, blazquez20}. Solid-phase reactions updated here are related to formic acid, ethanol, and acetaldehyde.

The second column of Table~\ref{table-mdls_ice_abun} (model B) shows the change in the gas-phase abundances of the species with the updated gas-phase reaction network. he inclusion of new gas-phase reactions has a minor impact on the chemistry, as the assumed warm-up timescale ($\thicksim$ 500 yr) is too short for these reactions to significantly influence the chemistry after ice species desorb from grains. Most of the volatile species have not come off the grains yet in the prestellar environment. These results indicate that if the recent explosion is the primary source of heating, the relative abundance ratios in AA2 are governed by ice chemistry and thermal desorption. However, it is important to note that this conclusion assumes that the recent explosion is the sole heating source. Given that the gas-phase formation of acetaldehyde via \ce{C2H5} + \ce{O} is efficient in warm environments~\citep{garrod22}, a series of prior heating events could have activated acetaldehyde formation, leading to a different chemical composition.

\begin{deluxetable}{ccccccc}
\tablewidth{0pt}
\tabletypesize{\footnotesize}
\tablecolumns{7}
\tablecaption{Ice abundances ($n(i)/n_\textrm{H}$) with the model update \label{table-mdls_ice_abun}}
\scriptsize
\tablehead{
\colhead{Model} & \colhead{A} & \colhead{B} & \colhead{C} & \colhead{D} & \colhead{E} & \colhead{F}  \\
\colhead{ } & \colhead{{\tt control}} & \colhead{A + gas reac} & \colhead{A + surf reac} & \colhead{B + C + high $A_\textrm{V,0}$} & \colhead{B + C + $B_\textrm{mag}$=0.3} & \colhead{D+E}
}
\startdata
FA & 1.4E-07 & 1.4E-07 & 1.0E-07 & 9.6E-08 & 1.6E-08 & 1.4E-08 \\
AA & 5.5E-09 &  5.5E-09 &1.1E-08 & 2.6E-08 & 1.1E-09 & 2.3E-08 \\
MF & 1.2E-07 & 1.2E-07 & 1.2E-07 & 1.1E-07 & 1.0E-07 & 1.1E-07 \\
GA & 3.0E-08 & 3.0E-08 & 3.0E-08 & 2.9E-08 & 2.6E-08 & 2.5E-08 \\
EtOH & 3.3E-07 & 3.3E-07 & 1.9E-07 & 5.2E-07 & 1.0E-08 & 2.0E-07 \\
\enddata
\end{deluxetable}

\begin{figure}
\centering
\includegraphics[width=0.6\textwidth]{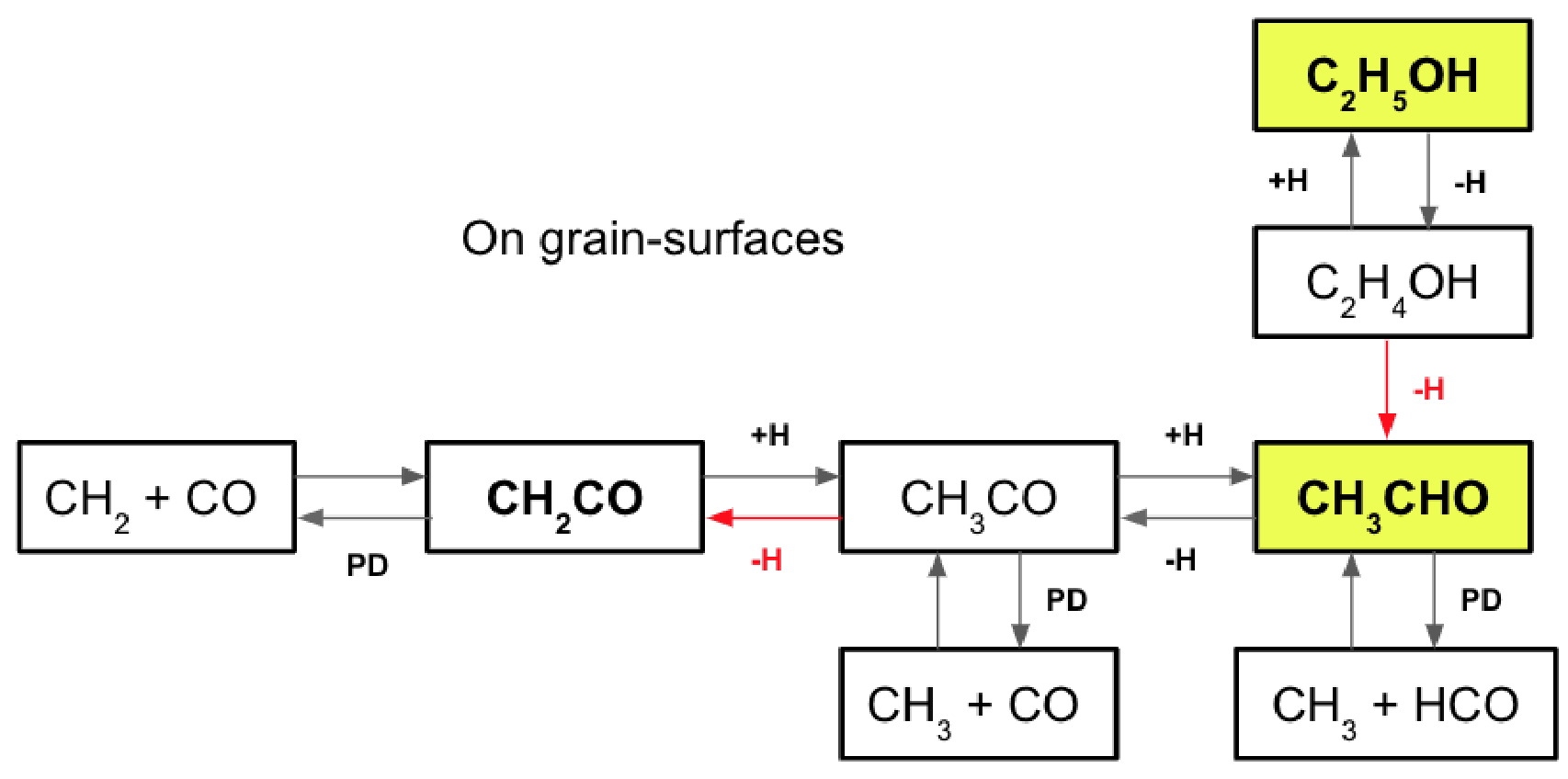}
\caption{Surface reaction diagram for the formation of acetaldehyde. The updated solid-phase reactions are denoted in red arrows.}
\label{figure-aa_rxns_dgrm}
\end{figure}

Table~\ref{table-mdls_ice_abun} (model C) summarizes the changes in the ice abundances at the end of the prestellar phase with the updated solid-phase reactions. All three species related to the network update (AA, FA, EtOH) show considerable changes. Adjusting the fraction for the CO + OH reaction to form COOH results in a 30$\%$ lower abundance of formic acid ice. The abundance of ethanol ice decreases by a factor of $\thicksim$ 2 due to the new fraction of the sink of \ce{C2H4OH} to form acetaldehyde. Interestingly, the inclusion of backward reactions for the hydrocarbon reaction chains does not change the abundances of icy ethanol much. Rather, the population of hydrocarbons has been rebalanced. The hydrogen and carbon budget was previously concentrated in the higher order of $\textrm{C}_{2}\textrm{H}_{n}$ such as \ce{C2H6}. However, with the new network, the abundances of the intermediate $\textrm{C}_{2}\textrm{H}_{n}$ increase, while the abundance of \ce{C2H6} decreases. This change can further affect the surface chemistry involving the intermediate hydrocarbon radicals. The abundance of acetaldehyde ice increases by a factor of $\thicksim$2 with the update of the surface network. Although the inclusion of H-abstraction of \ce{CH3CO} into the chemical network effectively reduces the abundance of acetaldehyde in the ice, the increase in the acetaldehyde ice from the new formation channel~ (H + \ce{C2H4OH}) dominates the decrease (see Figure~\ref{figure-aa_rxns_dgrm}). This efficient formation of acetaldehyde from the precursor of ethanol could explain the spatial coincidence between ethanol peak and AA2. Although there is some improvement in describing the observed ratios with the update of the chemical network, the modeled ratios are still far off from the observations. This implies that other explanations are needed to better understand the chemistry of AA2. 

\paragraph{collapse timescales and visual extinction} \label{subsubsec:phys_param}
The chemistry in our models of star-forming regions is mainly governed by three physical factors -- density, temperature, and visual extinction. By varying the model parameters related to them, we examined how the ice composition is affected by different physical conditions. As an experiment, we set a higher visual extinction model by adopting an initial visual extinction of $A_\textrm{V,0}$=3. The model with the updated chemical network is used as a basis. Table~\ref{table-mdls_ice_abun} compares the ice composition of the higher $A_\textrm{V,0}$ model (model D) with the {\tt control}. In the experiment~(higher $A_\textrm{V}$ environment), the ice abundances of ethanol and acetaldehyde increase  by a factor of two while the change in the other three species is minor. The increase in the abundances arises in the early stages of collapse. In such low-density environments, the dust temperature of the higher $A_\textrm{V}$ model is lower, because heating by the external UV photons is suppressed. This results in slower diffusion of atomic hydrogen on grain surfaces, which ultimately retards the driving of the chemical complexity. However, it should be noted that the destruction of large ice species via UV photodissociation (PD) is also suppressed in the high $A_\textrm{V}$ model. Thus, the new ice composition is determined by the balance between the slower formation of large ice species through H diffusion and the slower PD destruction of the species. Ethanol and acetaldehyde ice would be more vulnerable to PD, so the abundances of ice in the high $A_\textrm{V}$ model (low UV flux) increase despite the lower diffusion rate of H. As for the other three species, the two chemical processes balance well, resulting in little changes in the ice abundances.

\begin{figure}
\centering
\includegraphics[width=0.6\textwidth]{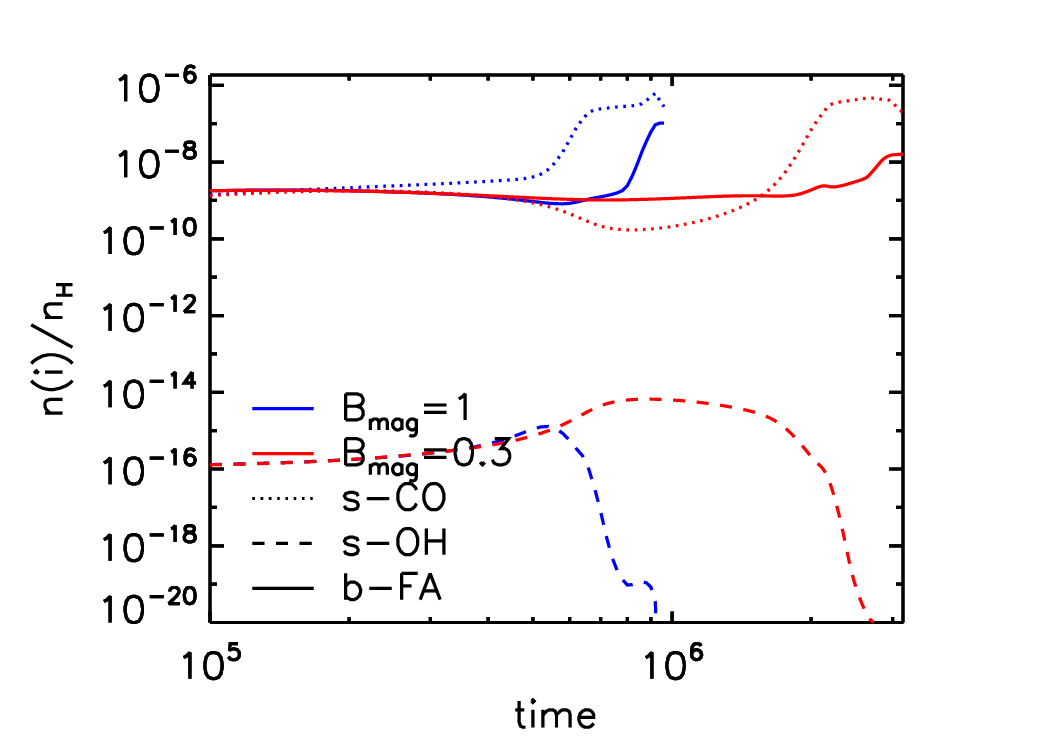}
\caption{Time-dependent ice abundances of CO and OH on the grain surface}
\label{figure-co_oh_timescale}
\end{figure}

The repercussion of a slower collapse timescale is examined in another experimental model E. The magnetic retardation factor for the collapse, $B_\textrm{mag}$, is adopted here to control the collapse timescale. This parameter takes a value between 0 (static) and 1 (free fall), and we set $B_\textrm{mag}=0.3$ for the experimental setup. This results in a collapse timescale approximately 3 times longer than the free-fall timescale. Table~\ref{table-mdls_ice_abun} (model E) summarizes the ice composition in the slower collapse model. Again, ice abundances of ethanol, acetaldehyde are significantly affected. The ice abundances of these species decrease by an order of magnitude. Considering that PD is a dominant destruction process of ethanol and acetaldehyde, the decrease in the abundances of these species can be explained by the longer exposure time to external UV photons due to slow density evolution. This result shows that ice abundances of acetaldehyde and ethanol are sensitive to the physical environment among the five COMs of investigation.

The ice abundance of formic acid also decreases by an order of magnitude when the density evolves slowly. Interestingly, the mechanism involved in this change differs from those of acetaldehyde and ethanol; It is more relevant to the timescale needed to form the reactants that produces formic acid. Figure~\ref{figure-co_oh_timescale} shows the abundances of two main building blocks (dashed and dotted lines) to form COOH (solid lines) on the grain surface over collapse. In both models, icy formic acid forms actively at the end of the evolution. It should be noted that the abundance of OH on the grain surface is so low that it may act as a bottleneck controlling the total rate of formation of surface COOH. In the free-fall collapse model~({\tt control}), the higher accretion rate of atomic O and H helps the supply of OH formed via the E-R mechanism (see $\sim8\times10^5$ yr). In the slower collapse model, the OH abundance on the surface (red dashed line) constantly decreases at late times of collapse, resulting in a lower ice abundance of formic acid.

Table~\ref{table-mdls_ice_abun} (model F) shows the ice abundances of the model adopting the higher initial visual extinction (model D) and the slower collapse timescale (model E). The abundance of acetaldehyde in the ice is similar to that of model D rather than that of model E. PD of acetaldehyde via external UV is suppressed at an early time despite the slow density evolution because of the high background visual extinction. This implies that the ice abundance of acetaldehyde can be high in well-shielded regions regardless of the dynamical history of early phase of star formation. The ice abundance of formic acid is insensitive to the initial background visual extinction, and it shows values similar to the slow collapse model. For ethanol, the impacts of slower collapse timescale and high visual extinction at the beginning are balanced out, and the abundance of ethanol is similar to that of model C.

The model F best reproduces the observational features and is used as a basis for further exploration of the desorption behavior of the species. It also should be noted that the ice abundance of this base prestellar model is similar to what is found toward HOPS 373SW (see Table~\ref{table-hops373sw}). HOPS 373SW is one of the protostellar systems located in the Orion molecular cloud, and the observation of the sublimated COMs by the recent burst of HOPS 373SW revealed its extremely young chemical characteristics~\citep{lee23}. Given that, it is expected that the chemical composition of HOPS 373SW should be similar to that of Orion KL. This similarity shows that the prestellar model F could be a good testbed for further study of the post-sublimation chemistry in Orion KL.

\begin{deluxetable}{ccc}
\tablewidth{0pt}
\tabletypesize{\footnotesize}
\tablecolumns{3}
\tablecaption{The relative abundances of COMs w.r.t. acetaldehyde ($n(i)/n$(\ce{CH3CHO})) in the gas-phase of HOPS 373SW and in the solid-phase of {F} model \label{table-hops373sw}}
\scriptsize
\tablehead{
\colhead{Species} & \colhead{HOPS 373SW}\tablenotemark{a} & \colhead{model F}
}
\startdata
MF & 8 & 5 \\
GA & 1 & 1 \\
EtOH & 3 & 9 \\
\enddata
\tablenotetext{a}{\citet{lee23}; Formic acid transition lines are not covered in the spws of this observation.}
\end{deluxetable}

\paragraph{Desorption energies} \label{subsubsec:e_des}
The relative abundance ratios between the species in the gas are significantly different from those in ice mantles; nevertheless, little chemical processing is expected after the ice species come off the grains because of the short warm-up timescale. This discrepancy between gas- and solid-phase chemistry is caused by the different thermal desorption behavior of the species.

The update of chemical networks and / or different physical parameters employed in the model does not fully explain the observed abundance ratios in the gas despite improvements in the description of ice compositions. In particular, the distinctively high abundance of ethanol and methyl formate compared to others in the observations (e.g. MF/FA$\thicksim$240) is hardly reproduced. The chemistry in the gas and on the grain-surface are coupled through the desorption / accretion process, and we could expect that the different treatment for the desorption of ice species would be necessary. Binding energies ($E_\textrm{des}$) are governing these processes within the model but these parameters have large uncertainties. In this section, a different set of $E_\textrm{des}$ is adopted based on the previous literature, and its impact on the estimated abundances in the gas is discussed. 

\begin{deluxetable}{cccc}
\tablewidth{0pt}
\tabletypesize{\footnotesize}
\tablecolumns{4}
\tablecaption{The update of binding energies of species \label{table-e_des}}
\scriptsize
\tablehead{
\colhead{Species} & \colhead{$E_\textrm{des, prev}$} &  \colhead{$E_\textrm{des, new}$} & \colhead{Notes}\\
\colhead{} & \colhead{(K)} & \colhead{(K)} & \colhead{}
}
\startdata
\ce{H2O} & 5700 & 4815 & \tablenotemark{a} \\
\ce{CH3CHO}; AA & 2775 & 4800 & \tablenotemark{b} \\
\ce{C2H5OH}; EtOH & 6259 & 5400 & \tablenotemark{b}  \\
\ce{HCOOCH3}; MF & 4210 & -- & \tablenotemark{c} \\
\ce{HCOOH}; FA & 5570 & -- & \tablenotemark{d} \\
\ce{CH2OHCHO}; GA & 5630 & -- & \tablenotemark{c} \\
\enddata
\tablenotetext{a}{\citet{sandfordandallamandola90}}
\tablenotetext{b}{\citet{wakelam17}}
\tablenotetext{c}{\citet{burke15}}
\tablenotetext{d}{\citet{collings04}}
\tablecomments{The new value for water is also consistent with the density functional theory (DFT) calculations by \citet{wakelam17}. The ethanol binding energy previously used in the model was estimated assuming a crystalline ice surface~\citep{lattelais11}. We used the lower ethanol binding energy based on recent DFT calculations for ASW~\citep{wakelam17}.}
\end{deluxetable}

Table~\ref{table-e_des} summarizes the updated set of $E_\textrm{des}$ under consideration. The lower binding energy for water $E_\textrm{des}$(\ce{H2O}) is newly assumed based on the experimental study for the surfaces of amorphous solid water (ASW) by \citet{sandfordandallamandola90}, although a wide range of values has been estimated for $E_\textrm{des}$(\ce{H2O}) depending on methodology and/or experimental setup~\citep[e.g.][]{wakelam17, fraser01}. The binding energy of acetaldehyde in the new model has changed because the previous value was significantly lower compared to the estimation from DFT calculation by~\citet[][5400~K]{wakelam17}. A recent experimental and theoretical study by \citet{ferrero22} predicted that acetaldehyde is more volatile than water. To meet this prediction, we adopted a slightly lower value (4800~K) than the binding energy of water. 

\begin{deluxetable}{ccccccccc}
\tablewidth{0pt}
\tabletypesize{\footnotesize}
\tablecolumns{9}
\tablecaption{Gas-phase abundances at different warm-up temperature \label{table-e_des_model_results-abun}}
\scriptsize
\tablehead{
\colhead{species}  & \colhead{observations} & \colhead{100 K} & \colhead{110 K} & \colhead{115 K} & \colhead{120 K} & \colhead{130 K} & \colhead{140 K} & \colhead{150 K} 
}
\startdata
FA	&	3.0E-10 &   1.7E-14	&	1.1E-11	&	1.2E-10	&	8.0E-10	&	2.8E-09	&	3.7E-09	&	5.9E-09	\\
AA	&	9.0E-10 &   9.8E-11	&	1.0E-09	&	1.7E-09	&	2.8E-09	&	5.7E-09	&	7.6E-09	&	1.2E-08	\\
MF	&	6.7E-08 &   7.6E-09	&	1.3E-08	&	1.5E-08	&	2.0E-08	&	3.3E-08	&	4.1E-08	&	6.0E-08	\\
GA	&$<$1.0E-10 &   2.1E-17	&	5.8E-15	&	2.9E-13	&	5.4E-11	&	2.8E-09	&	5.2E-09	&	9.6E-09	\\
EtOH&	1.8E-08 &   2.0E-12	&	7.9E-10	&	5.2E-09	&	1.7E-08	&	4.2E-08	&	5.5E-08	&	8.7E-08	\\
\enddata
\end{deluxetable}

\begin{deluxetable}{ccccccccc}
\tablewidth{0pt}
\tabletypesize{\footnotesize}
\tablecolumns{9}
\tablecaption{Relative abundance ratios w.r.t. FA column densities at different warm-up temperature \label{table-e_des_model_results-ratio}}
\scriptsize
\tablehead{
\colhead{} & {observation} & \colhead{100 K} & \colhead{110 K} & \colhead{115 K} & \colhead{120 K} & \colhead{130 K} & \colhead{140 K} & \colhead{150 K} 
}
\startdata
AA	&	3 & $>$1000	&	92   	&	14	&	4	&	2	&	2	&	2	\\
MF	&	240 & $>$1000 & $>$1000 	&	125	&	25	&	12	&	11	&	10	\\
GA	&	$<$0.3 & $<$0.1	&	$<$0.1	&	$<$0.1	&	0.1	&	1.0	&	1.4	&	1.6	\\
EtOH	&	64 & 117	&	71	&	42	&	22	&	15	&	15	&	15	\\
\enddata
\end{deluxetable}

The models with this new set of binding energies are tested with a range of $T_\textrm{kin,max}$ from 100 K to 150 K. Table~\ref{table-e_des_model_results-abun}-\ref{table-e_des_model_results-ratio} shows the new abundances of species in the gas and their abundance ratios with respect to (w.r.t.) a formic acid column density, respectively. Note that the modeled chemical compositions change sensitively in response to temperature, particularly for acetaldehyde and methyl formate. This high variability of the ratios for these two species seen in Table~\ref{table-e_des_model_results-ratio} is related to their high volatility compared to those of other species. The lower water binding energy in the new model is related to this. Water is the most abundant constituent of ice, and the change in $E_\textrm{des}$(\ce{H2O}) can affect the general desorption characteristics within the model. In particular, ice species with binding energies lower than $E_\textrm{des}$(\ce{H2O}) more easily co-desorb with water; otherwise, ice species desorb relatively slowly until high enough thermal energies are provided. As the water binding energy is adjusted to the lower value in the new model, methyl formate and acetaldehyde become the only species whose binding energies are smaller than water among the five ice species under investigation, achieving high volatility compared to others. The high abundance ratio of methyl formate as a result of this distinctive characteristic well explains its observational features, such as extended morphologies and high column densities.

The observed chemical composition is adequately consistent with the model results at $T_\textrm{kin,max}$=115 K. It should be noted that the observed ratios are not necessarily well-represented by a single chemical model. This is because the chemical model represents the chemistry at one single density point while the observations cover a finite spatial scale integrated throughout the line of sight. Considering the high variability of modeled chemical compositions depending on temperature, the observed ratios can be interpreted as a result of integration of chemistry with a temperature gradient.

\subsubsection{Implication from the chemical model} \label{subsubsec:aa_chem}
If the warm-up timescale of HC-SW is as short as $\sim$ 500 years due to the heating by a recent explosive event~\citep{zapata11}, the observed relative abundance ratios between aldehyde-like species would be determined by the ice composition as a result of ice chemistry and thermal desorption. The observed chemistry toward AA2 is the result of the integration of chemistry throughout the line of sight with a temperature gradient. The lower binding energy of methyl formate and acetaldehyde compared to that of water and other species is needed for a reasonable description of the desorption behavior of the species depending on the temperature.

The main pathways for the formation / destruction of acetaldehyde on the surface of the grain are described in Figure~\ref{figure-aa_rxns_dgrm}. The net sources of this species are radical-radical recombination (\ce{CH3} + HCO) and nondiffusive reactions between CO and radicals (\ce{CH2} + CO / \ce{CH3} + CO) followed by hydrogenation~\citep{jinandgarrod20}. Because the thermal diffusion of heavy radicals on grain surfaces is inefficient in prestellar environments, the latter would predominantly contribute to the formation of acetaldehyde. In addition to that, the H-abstraction reaction of the precursor of ethanol (H + $\ce{C2H4OH}$) considered here would be another important pathway for the formation of acetaldehyde. This explains well the spatial coincidence between EtP and AA2. This emphasizes the direct surface chemical link that involves atomic hydrogen for the two species, as suggested by \citet{bisschop07} and \citet{fedoseev22}. However, in the chemical model, the top-down surface chemistry, where the destruction of ethanol forms acetaldehyde, is more important because of the higher ice abundance of ethanol. The H-diffusion reactions in this chemical network do not contribute to the net formation / destruction of acetaldehyde, but rather involve the chemical cycle between acetaldehyde and its precursors. Recurrent formation of acetaldehyde through hydrogenation is an important mechanism by which acetaldehyde ice is chemically desorbable at low temperature. As discussed in \S~\ref{subsubsec:phys_param}, the destruction of acetaldehyde ice is dominated by PD, showing the high abundance of ice in environments with low UV flux. This vulnerability of acetaldehyde ice to PD could explain the common in the detection of acetaldehyde in quiescent environments in the literature.

It should be noted that the models in this study are constructed under the assumption that the ``hot core" is actually a rapidly heated pre-existing density enhancement~\citep{zapata11}. We assumed that this core had formed in quiescent environments because of the uncertainties of the pre-explosion environments. However, given the complex environments of Orion KL, the "hot core" may undergo energetic events multiple times. Modeling the chemical evolution with a series of heating events could be relevant in this case.

\section{discussion} \label{sec:discussion}
\subsection{Spatial distribution acetaldehyde and other HCO-bearing species} \label{subsec:discuss-spat_distribution}

The spatial distribution of acetaldehyde is compared with \ce{C2H5OH} (ethanol; EtOH) and other HCO-bearing species from TCL18. TCL18 found that ether (the species containing C-O-C) generally exhibit an extended V-shaped morphology linking HC, CR, and BN objects, with the brightest peak toward CR. In contrast, the species with a hydroxyl (C-O-H) group show weaker or missing emission in CR, with their brightest peaks associated with HC-SW.

Acetaldehyde exhibits bright emission near HC-SW and weak or no emission toward \mj{the} CR, consistent with the trend observed in TCL18. Notably, this peak near HC-SW (AA2) coincides with the ethanol peak in TCL18, suggesting a potential chemical correlation between acetaldehyde and ethanol. Given that this peak is also a source of other COMs, it is likely linked to the sublimation of ice. However, the strongest emission of acetaldehyde is found in the northern part of the hot core, where other O-bearing COMs are weak. Interestingly, while its isomer, \ce{c-C2H4O} shares a similar spatial distribution, it does not peak at AA1~(Liu, Priv. Comm.), suggesting unique aldehyde-forming chemistry at AA1. One possible explanation is that a series of prior heating events triggered gas-phase reactions leading to acetaldehyde formation. In our chemical model, a single brief warm-up is unlikely to significantly impact its abundance in the gas; however, repeated heating could enhance its formation. Notably, other O-bearing COMs are faint in this region, whereas acetaldehyde and N-bearing COMs, such as \ce{CH3CN and C2H3CN}, are bright. These species are known to have efficient gas-phase formation pathways~\citep{garrod22}.

The potential chemical correlation between acetaldehyde and ethanol is further supported by experimental studies. \citet{bisschop07} and \citet{fedoseev22} demonstrated that atomic H bombardment on acetaldehyde ice can lead to ethanol formation via hydrogenation. Conversely, \citet{skouteris18} proposed a top-down gas-phase pathway where the ethanol destruction produces acetaldehyde. Our chemical modeling suggests a top-down ice-phase process instead, where acetaldehyde forms through hydrogen abstraction from ethanol. This aligns with recent quantum chemical calculations, which indicate that acetaldehyde is considerably resistant to direct hydrogenation to form ethanol in icy environments~\citep[]{molpeceres25}.

\begin{figure*}
\gridline{\fig{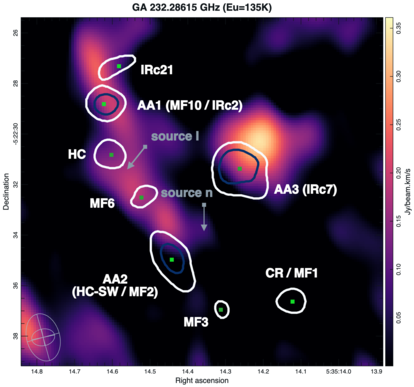}{0.5\textwidth}{}
          \fig{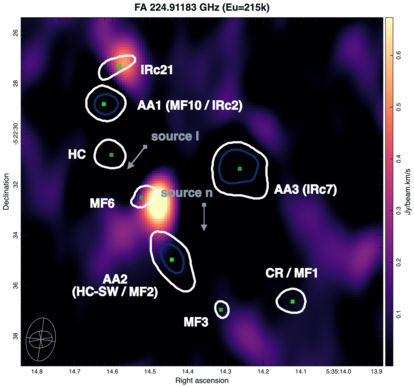}{0.5\textwidth}{}}
   \caption{Moment 0 maps of glycolaldehyde (left) and formic acid (right) generated by integrating the velocity channel from $v_\textrm{LSR}=0.0$ km/s to $v_\textrm{LSR}=10.5$ km/s. The synthesized beam is shown in the bottom left corner. The contour of the high-energy transition of acetaldehyde is overlaid. The GA and FA maps are generated by using the same transition lines and data set reported in TCL18. As for the formic acid map, the higher energy transition showed a very similar distribution. The synthesized beams for the acetaldehyde map ($0.68\arcsec\times0.50\arcsec$) and SV data ($1.71\arcsec\times1.15\arcsec$) are shown in the bottom left corner. Velocity vectors of the proper motion of source I and n are also denoted with gray arrows.}
\label{figure-mom0_ga_fa}
\end{figure*}

The spatial distribution of acetaldehyde differs from that of another aforementioned aldehyde, glycolaldehyde, as shown in the left panel of Figure~\ref{figure-mom0_ga_fa} and \ref{appendix:mom0_ga}. While both species exhibit bright emission near HC-SW and weak or no emission toward CR, glycolaldehyde emission in HC-SW is confined closer to the ethylene glycol peak (see also Figure 1 in \citet{tercero18}). Furthermore, glycolaldehyde emission is not detected in AA1\footnote{Despite the presence of a signal near AA1 seen in the left panel of Figure~\ref{figure-mom0_ga_fa}, the detection of glycolaldehyde in this region is not claimed, as emissions from other transition lines are absent. For instance, the moment 0 map of the 223.11525 GHz glycolaldehyde transition line, where no emission from AA1 emission is detected, is shown in figure~\ref{appendix:mom0_ga}.}.

This lack of spatial correlation between glycolaldehyde and acetaldehyde suggests that the dominant COM chemistry in this region is not dictated solely by the sublimation of ice but may also be influenced by Orion KL's complex physical environment. Alternatively, if early ice compositions are imprinted in shaping the observed chemical distribution, the spatial segregation of these species may not be governed solely by their shared building blocks. One possible explanation is the widespread presence of HCO radicals across the region -- if HCO radicals are abundant on grain surfaces regardless of local conditions, the availability of the other reacting radical may act as the bottleneck that determines chemical segregation. Additionally, differences in binding energies and the excitation behavior of molecules may play a more crucial role in shaping their spatial distributions.

However, we lean toward attributing the lack of spatial correlation to the complex physical environment of Orion KL. This is because some statistical studies have found strong correlations between acetaldehyde and glycolaldehyde in other regions. For example, \citet{baek22} examined twelve high-mass star-forming regions using ALMA and reported a strong correlation between the column densities of these two species. Furthermore, except for AA1 -- where a different thermal history is expected -- the overall distributions of acetaldehyde and glycolaldehyde are coincident. The offset between their peaks at HC-SW may be attributed to differences in their binding energies. This suggests that a chemical link driven by their shared functional group, imprinted during early ice formation, can emerge depending on the source. These findings emphasize the need for more observations across diverse spatial scales and larger sample sizes to better understand chemical correlations based on functional groups.

The spatial distribution of formic acid and acetaldehyde is presented in the right panel of Figure~\ref{figure-mom0_ga_fa}. Similar to glycolaldehyde, formic acid emission does not strongly correlate with acetaldehyde. However, two compact peaks of formic acid are located in the southwestern and northern regions (IRc21) of the HC, where weak acetaldehyde emission appears. The formic acid emission in HC-SW aligns closely with other O-bearing COMs reported in TCL18, suggesting that it likely originates from the sublimation of icy grain mantles. In contrast, the co-presence of formic acid and acetaldehyde in IRc21 may have a different origin; In addition to acetaldehyde investigated in this study, formic acid is the only species that exhibits emission peaks toward IRc21 among the fourteen O-bearing COMs investigated in TCL18. Acids generally have higher binding energies, while acetaldehyde is more volatile~\citep{ferrero22} (See the right panel of Figure~\ref{figure-mom0_ga_fa} in this study for acetaldehyde and Figure 1 in TCL18 for others). Therefore, if both species originated from ice mantle sublimation, other O-bearing COMs would be expected, but this is not observed.

An alternative explanation is ``soot line" chemistry in the gas. This region is likely associated with high-energy-density gas, as evidenced by vibrationally excited \ce{HC3N} transition lines detected in this region~\citep{peng17}. Additionally, bright lines of N-bearing COMs are observed in this region~\citep[see Figure 2 in ][]{pagani17}. At temperatures around 300~K, the sublimation of carbon grain occurs, producing large fragments. These large species are broken down in the hot gas, producing an excess of hydrocarbons, nitriles with high excitation temperature, and a deficiency in O-bearing COMs~\citep{vanthoff20}. This signature chemical composition is likely consistent with what is observed toward IRc21. If this is the case, formic acid and acetaldehyde in IRc21 could be secondary-generation species formed in the gas. \citet{garrod22} demonstrated that a major production pathway for acetaldehyde exists in the gas phase once the ice mantles have fully desorbed. Additionally, \citet{bbh99} showed that one of the primary products of the gas-phase reaction between OH-radicals and benzene is formic acid.

As seen in the Figure 1 of TCL18, methyl formate shows a different morphology from acetaldehyde. Methyl formate emission follows a typical V-shaped distribution of ether. The weak acetaldehyde emission toward CR is consistent with the proposition of TCL18 in that acetaldehyde does not have a C-O-C functional group. To explain this chemical segregation, TCL18 proposed two scenarios that are not necessarily exclusive of each other. One is based on grain surface chemistry, that different radicals, methoxy (\ce{-CH3O}), and hydroxymethyl (\ce{-CH2OH}) dominate the radical-radical surface reactions of CR and HC-SW, respectively, driving different chemical complexity in the region. The other is the ion-neutral reaction in the gas of CR, where alcohol (mainly methanol) is actively destroyed to form C-O-C-bearing COMs. However, the bright acetaldehyde emission toward HC-SW despite the absence of C-O-H structure within the species indicates that either methoxy or hydroxymethyl radicals are not solely contributing to the chemical segregation toward this region. Given the higher excitation temperature of HC than CR \citep[$\Delta T\ga50$~K at high angular observations]{favre11, wilkins22}, different desorption behaviors of COMs depending on binding energies can contribute to the spatial distribution of COMs.

In summary, a strong emission of acetaldehyde is observed in both northern and southwestern regions of the hot core. The acetaldehyde peak at HC-SW is likely linked to the sublimation of icy grain mantles, alongside other HCO-bearing COMs, while the northern peak suggests a distinct aldehyde-forming chemistry, potentially due to a different thermal history. Despite its likely ice origin in HC-SW, there is an offset between acetaldehyde peak and those of HCO-bearing species such as glycolaldehyde and formic acid. This discrepancy may arise from differences in binding energies, excitation behavior, or the ubiquitous characteristic of the HCO radical. Consequently, no clear chemical segregation is observed for HCO-bearing species across the region.

\subsection{Searching for other aldehyde species} \label{subsec:discuss-other_aldehydes}
Propynal ($\ce{C2HCHO}$), propenal ($\ce{C2H3CHO}$), and propanal ($\ce{C2H5CHO}$) are the aldehydes with three carbon atoms derived from propyne, propene, and propane. They make up three of the five large ($n_\textrm{atom} \ge 6$) aldehydes that have been detected in molecular clouds~\citep{mcguire18b}. Only tentative detections for propanal and propoenal have been reported toward Orion KL using ALMA SV data~\citep{pagani17}. Motivated by this, we searched for the three C$_{2}$H$_{n}$CHO species from our new archive data, following the same analysis in \S~\ref{subsec:results-spectra}. However, this attempt was not successful because the number of bright transitions of C$_{2}$H$_{n}$CHO species in the spectral window was low, and all the investigated transitions did not meet the detection criterion. The non-detection of C$_{2}$H$_{n}$CHO in Orion KL can be explained by the combination of their weak line properties and/or low abundances of the species.

After the first detection of emission toward TMC-1~\citep[propynal only]{irvine88} and of absorption toward Sgr B2 (N)~\citep{hollis04}, these three C$_{2}$H$_{n}$CHO have been investigated toward other interstellar environments. Although a single detection of warm propanal ($\thicksim$125 K) has been reported for IRAS16293-2422 with ALMA~\citep{lykke17}, successful detections of these species were mostly made toward cold components~\citep{irvine88, hollis04, requena-torres08, loison16}. This temperature-sensitive detection of C$_{2}$H$_{n}$CHO may indicate that the formation of C$_{2}$H$_{n}$CHO is favored in cold (10-20 K) environments. Further investigation of C$_{2}$H$_{n}$CHO combined with a chemical model would be interesting to verify this proposition. Laboratory and quantum-mechanical studies for these aldehydes would be necessary to keep pace with recent astronomical discoveries and to construct a robust chemical network for these species.

\section{Summary} \label{sec:summary}
Thanks to the significant improvement in the sensitivity and angular resolution of ALMA, we can report the distinct detection of acetaldehyde toward Orion KL for the first time. The main conclusions of our study are enumerated below:

\begin{enumerate}
\item Three acetaldehyde emission components are identified with 10$\sigma$ confidence, and five additional weaker peaks are observed at 5$\sigma$. These components follow the general morphologies of non-ether COMs; the primary emission peak (AA2) is associated with HC-SW, while the emission toward the chemically rich region CR is faint.

\item A strong emission of acetaldehyde is observed in both the northern and southwestern regions of the hot core. The acetaldehyde peak at HC-SW (AA2), is likely linked to the sublimation of icy grain mantles, along with other HCO-bearing COMs. The northern peak (AA1) suggests a distinct aldehyde-forming chemistry, potentially influenced by a different thermal history.

\item The acetaldehyde peak exactly matches the ethanol peak in HC-SW. According to the chemical model, a top-down surface chemistry process, where the destruction of ethanol forms acetaldehyde, may play a key role in this correlation.

\item The spatial distributions of acetaldehyde are compared with those of other HCO-bearing species such as glycolaldehyde and formic acid. The emission from these species is absent around AA1, reflecting the distinct origin of acetaldehyde in this region. Despite the likely ice origin in HC-SW, an offset exists between the acetaldehyde peak and those of other HCO-bearing species. This discrepancy may arise from differences in binding energies, excitation behavior, or the ubiquitous characteristic of the HCO radical. Consequently, no clear chemical segregation is observed for HCO-bearing species across the region.

\item Acetaldehyde column densities are estimated using LTE line modeling toward AA1 and AA2. For AA2, the results are compared with other HCO-bearing species and ethanol from the literature. The relative abundance ratios between the species are estimated to be GA:FA:AA:EtOH:MF =~$<$0.3:1:3:60:240. 

\item Chemical modeling of AA2 is performed to explain the observed abundance ratios toward AA2. The contribution of gas-phase chemistry in this model was found to be minor because of the very short warm-up timescale. The ice composition and desorption behavior of the molecules of interest critically determine the observed ratios. To explain the observed abundance ratios, the desorption energy of water should be higher than that of methyl formate but lower than other species under consideration. 

\item The larger aldehyde species with three carbon atoms (C$_{2}$H$_{n}$CHO) are not detected in this investigation. This possibly implies that the formation of C$_{2}$H$_{n}$CHO is favored in cold (10-20 K) environments.

\end{enumerate}

\section*{Acknowledgements}
MJ, SBC and MAC were supported by the Goddard Center for Astrobiology and by the NASA Planetary Science Division Internal Scientist Funding Program through the Fundamental Laboratory Research work package (FLaRe). R.T.G. thanks the National Science Foundation for funding through the Astronomy \& Astrophysics program (grant number 2206516). GB was supported by Basic Science Research Program through the National Research Foundation of Korea(NRF) funded by the Ministry of Education(grant number RS-2023-00247790). This paper makes use of the following ALMA data: ADS/JAO.ALMA\#2011.0.00009.S, ADS/JAO.ALMA\#2017.1.01149.S, ADS/JAO.ALMA\#2016.1.00297.S. ALMA is a partnership of ESO (representing its member states), NSF (USA) and NINS (Japan), together with NRC (Canada), MOST and ASIAA (Taiwan), and KASI (Republic of Korea), in cooperation with the Republic of Chile. The Joint ALMA Observatory is operated by ESO, AUI/NRAO and NAOJ. The National Radio Astronomy Observatory and Green Bank Observatory are facilities of the U.S. National Science Foundation operated under cooperative agreement by Associated Universities, Inc. We sincerely thank S.-Y. Liu and G. Molpeceres for the insightful comments and valuable discussion that contributed to this work.

\appendix
\renewcommand\thefigure{A.1}

\begin{figure}
\centering
\includegraphics[width=0.8\textwidth]{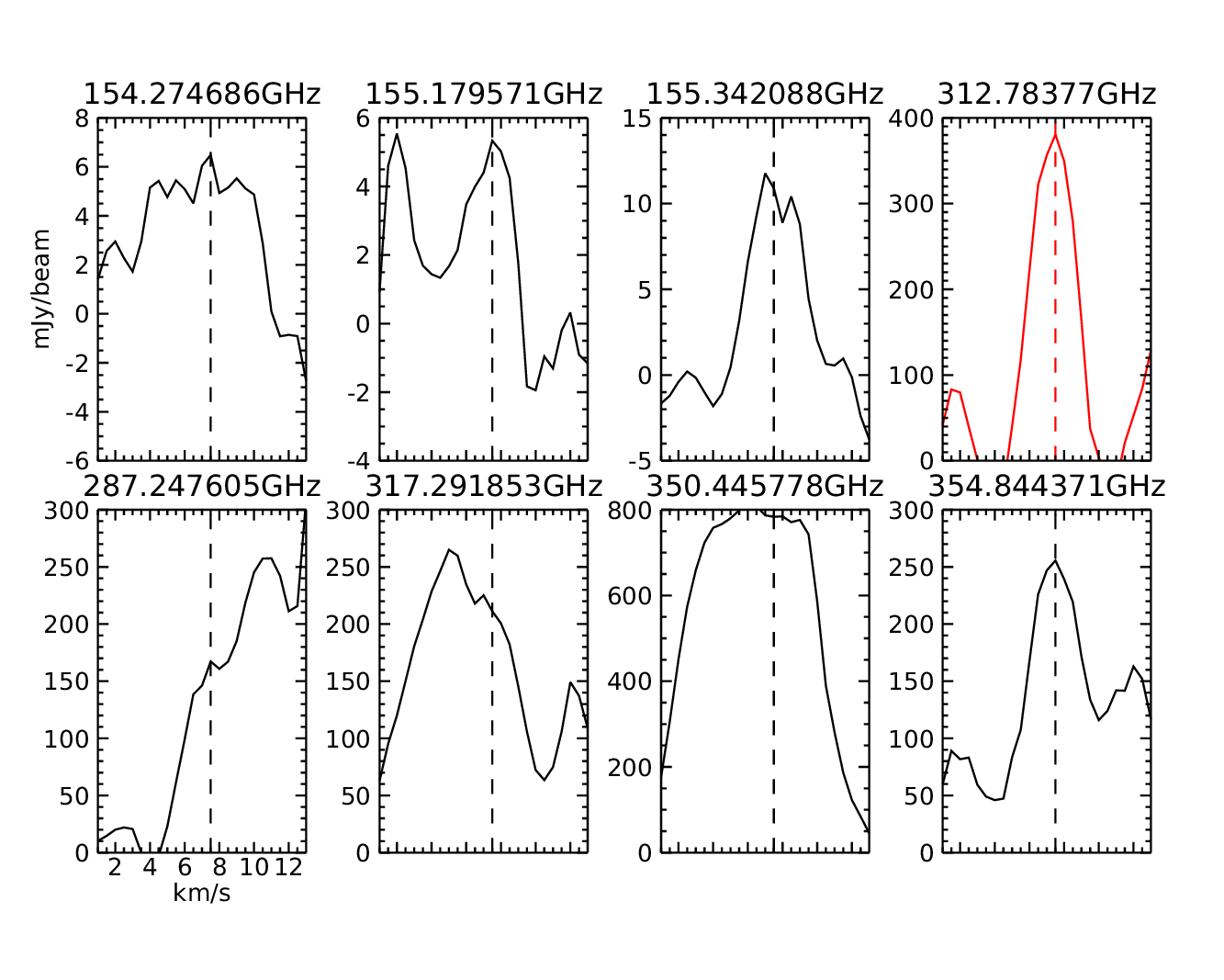}
\caption{The black lines represent the seven transitions extracted from AA1, that are predicted to be detectable based on LTE line modeling, but were excluded from further analysis due to severe blending. For comparison, the line profile of the representative transition at 312.78377 GHz is also shown in red in the top-right panel.}
\label{figure-a1}
\end{figure}

\section{The detailed description of the chemical model} \label{appendix:magickal}
Notably, the chemical model used here includes not only the typical diffusive grain-surface/ice chemistry, but also a range of nondiffusive mechanisms, by which reactants on the grain/ice surface or in the bulk ice may meet each other and react~\citep[see Appendix~\ref{appendix:nondiffusive};][]{jinandgarrod20, garrod22}. Another major characteristic of this version of {\em MAGICKAL} code is the elimination of bulk diffusion for all species but atomic and molecular hydrogen. Much of the chemistry within the bulk is thus driven either directly by photodissociation (via the PDI mechanism), or indirectly, as the result of the diffusion of H-atom and its reaction with existing bulk species. Modified rate equations are used to simulate the stochastic behavior of the grain-surface chemistry, where required \citep{garrod08b}. 
We add or update the reactions related to the HCO-bearing species and ethanol based on the chemical network of \citet{garrod22}. The base network includes a selection of new methylene- and methylidyne-related reactions on the grains, a more complete glycolaldehyde-related chemistry, and some key gas-phase reactions for COMs, such as the reaction of atomic O and the radical CH$_3$OCH$_2$ to produce MF \citep{Balucani15}, and the radiative association of CH$_3$ and CH$_3$O radicals to form dimethyl ether \citep{Tennis21}.

\section{The derivation of rate coefficient for hydrogen abstraction reaction of methyl formate} \label{appendix:kmf}
The gas-phase reactions newly included here are the neutral-neutral reactions where the four species of interest (MF, FA, AA, and EtOH) are destroyed through H-abstraction by OH. Differently from ethanol and acetaldehyde, the rate coefficient for methyl formate is not parameterized, so the experimental results are not directly applicable to the chemical model. Instead, we extrapolate the measurement from two different ranges of temperature (22-64 K; Jim{\'e}nez et al. 2016 and 233-372 K; Le Calv{\'e} et al. 1997), and adopt the constant $k_\textrm{OH+MF}$=5$\times10^{-13}~\textrm{cm}^{3}$/s. This roughly represents the rate coefficient at 120 K $<T<$ 150 K, which is consistent with the range of temperature assumed in the warm-up phase of the chemical model~(see Figure~\ref{figure-k_mf}). The neutral-neutral reaction for formic acid and glycolaldehyde has not been measured at low temperature. We adopted the same $k_\textrm{MF}$ for these two species, because the rate coefficient measured at $T>$ 200 K behaves similarly to methyl formate. Taking into account the high $k_\textrm{MF}$ measurement at $T~<$ 60~K (figure~\ref{figure-k_mf}), assuming the constant value $k$ for a wide temperature range can result in a significant underestimation of the impact of reactions in cold environments. To examine this, higher $k$ values are also tested in the prestellar model. It is found that the new gas-phase reactions little affect the chemistry regardless of the assumed reaction rate coefficients. This is because most of the volatile species have not come off the grains yet in the prestellar environment. 

\renewcommand\thefigure{A.2}
\begin{figure}
\centering
\includegraphics[width=0.6\textwidth]{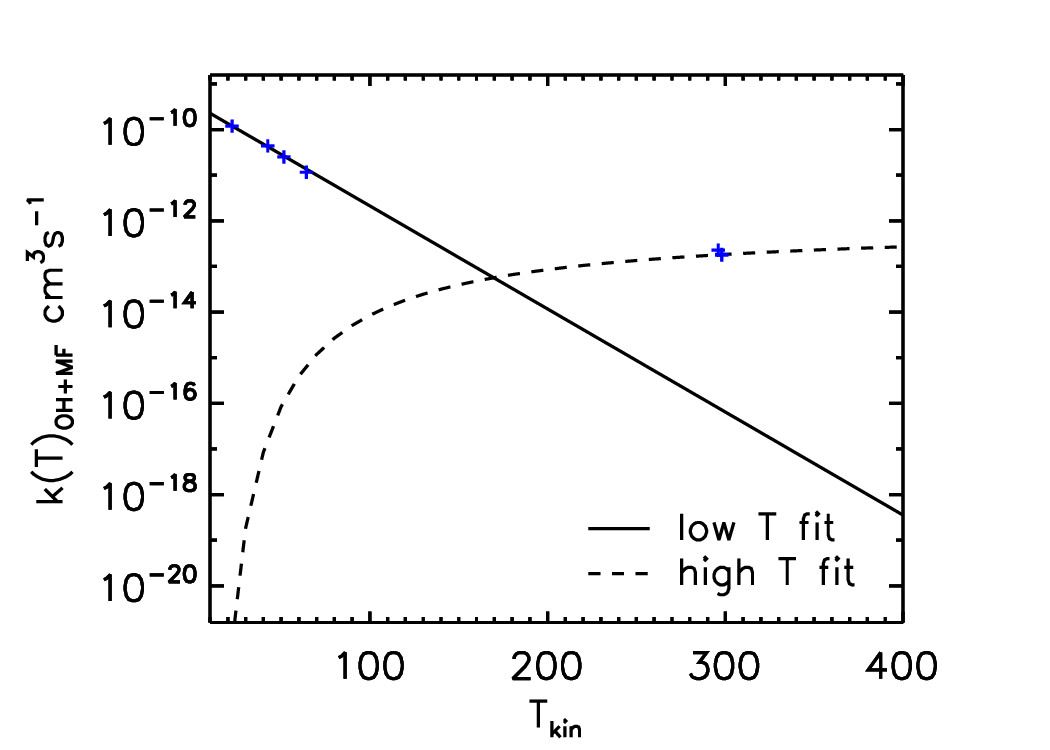}
\caption{The extrapolation of $k$ for methyl formate is based on two different experimental datasets. The experimental data and the parameters used to fit the temperature dependence of $k_\textrm{OH+MF}$ are adopted from \citet{jimenez16} and \citet{lecalve97}.}
\label{figure-k_mf}
\end{figure}

\section{Update of the chemical network in the ice} \label{appendix:solid_network}
For formic acid, two key reactions on grain surfaces primarily contribute to its formation:
\begin{eqnarray}
\rm HCO~+~OH \rightarrow \rm HCOOH \\
\rm CO~+~OH \rightarrow \rm \ce{CO2}~+~H~/~COOH.
\end{eqnarray}
In the second reaction, a small fraction of CO + OH reaction products can form COOH, which may then be hydrogenated to form formic acid. The branching ratio (BR) between the formation of COOH and \ce{CO2} + H has not been well-constrained, so in the {\tt control} model, we assumed that 1$\%$ of the CO + OH reaction products would form COOH. However, the high ice abundance of formic acid observed in the {\tt control} suggests that this BR might be too high. To address this, we empirically adjusted the BR in the new network, lowering it to 0.1$\%$ (a factor of ten lower).

The update of the solid-phase reactions focuses particularly on ethanol. In the {\tt control}, two primary reactions contribute to the ethanol formation: 
\begin{eqnarray}
\rm CH_{3}OH~+~CH_{2} \rightarrow \rm C_{2}H_{5}OH \\
\rm C_{2}H_{4}OH/C_{2}H_{5}O~+~\ce{H} \rightarrow \rm C_{2}H_{5}OH
\end{eqnarray}
The radicals \ce{C2H4OH} and \ce{C2H5O} in the second reaction are produced through ongoing hydrogenation of hydrocarbon in the model; \ce{C2H} -- \ce{C2H2} -- \ce{C2H3} -- \ce{C2H4} -- \ce{C2H5} -- \ce{C2H6}. We discovered that only the forward reactions were previously considered, leading to a potential overproduction of ethanol in the ice phase. To correct this, we have now included the backward reactions in this chain.

The surface reaction network of acetaldehyde is shown in Figure~\ref{figure-aa_rxns_dgrm}. Two newly included hydrogen abstraction reactions, highlighted by red arrow, contribute to acetaldehyde formation. A small fraction of H + \ce{C2H4OH} reaction products can now produce acetaldehyde, offering a competitive pathway against ethanol formation. This addition affects both the ethanol and acetaldehyde abundances in the ice. Additionally, the hydrogen abstraction of the \ce{CH3CO} radical is now considered to balance the hydrogenation of \ce{CH2CO}.

\section{nondiffusive mechanisms} \label{appendix:nondiffusive}
The new nondiffusive mechanisms considered in the model include three-body (3-B) reactions, photodissociation-induced (PDI) reactions, and the Eley-Rideal process. In the 3-B reactions, a surface or bulk reaction is immediately followed by the reaction of the product with a nearby species. In PDI reactions, a radical produced by photodissociation meets its reaction partner in the immediate vicinity right after dissociation. The Eley-Rideal process involves gas-phase species directly absorbing onto a grain-surface reaction partner.

The model incorporated the updated ``PDI2'' treatment of \citet{garrod22}, which allows for the recombination of photodissociation products in the bulk ice when no immediate reaction partner is available. Additionally, \citet{jinandgarrod20} proposed a three-body excited-formation (3-BEF) mechanism, where chemical energy produced by an initiating reaction in the 3-B process enables the overcoming of activation energy barriers. This mechanism has been shown to be important for the grain-surface production of methyl formate, helping explain its presence in the gas phase in cold, prestellar environments. The generic 3-BEF method provided by \citet{garrod22} is used here, which statistical treats the chemical energy available from the preceding reaction to overcome the barrier for a specific subsequent reaction.

\renewcommand\thefigure{A.3}
\begin{figure}
\centering
\includegraphics[width=0.6\textwidth]{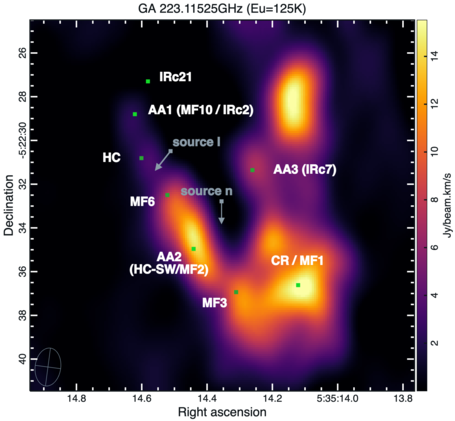}
\caption{The Moment 0 map for the 223.11525 GHz transition line of glycolaldehyde was generated by integrating the velocity channel from $v_\textrm{LSR}=0.0$ km/s to $v_\textrm{LSR}=10.5$ km/s. Notably, the morphology differs from that of 232.28615 GHz line seen in Figure~\ref{figure-mom0_ga_fa}, due to the presence of nearby methyl formate line at 223.11331 GHz. As illustrated in this figure, no glycolaldehyde emission is detected, despite its similar $E_\textrm{up}$ to the 232.28615 GHz line. Furthermore, TCL18 also reported the absence of glycolaldehyde emission at the 242.95781 GHz line ($E_\textrm{up}=148$ K) toward AA1. This suggests that the emission observed at AA1 the 232.28615 GHz line toward AA1 may not be real.}
\label{appendix:mom0_ga}
\end{figure}

\end{document}